\documentclass[usenatbib]{mnras}
\usepackage{hyperref}
\hypersetup{draft}
\usepackage{newtxtext,newtxmath}
\usepackage[T1]{fontenc}
\usepackage{ae,aecompl}
\usepackage{macros}

\usepackage{geometry}
\usepackage{fancyhdr}
\usepackage{amsmath}
\usepackage{graphicx} 
\usepackage[caption=false]{subfig}
\usepackage{lastpage}
\usepackage{etoolbox}
\usepackage{threeparttable}

\title[Formation of MW-mass Galaxies in FIRE]{The Formation Times and Building Blocks of Milky Way-mass Galaxies in the FIRE Simulations}
\author[Santistevan et al.]{
Isaiah B. Santistevan$^{1}$\thanks{E-mail: ibsantistevan@ucdavis.edu},
Andrew Wetzel$^{1}$,
Kareem El-Badry$^{2}$,
Joss Bland-Hawthorn$^{3,4}$,
\newauthor
Michael Boylan-Kolchin$^{5}$,
Jeremy Bailin$^{6}$,
Claude-Andr{\'e} Faucher-Gigu{\`e}re$^{7}$,
\newauthor
Samantha Benincasa$^{1}$
\\
$^{1}$Department of Physics \& Astronomy, University of California, Davis, CA 95616, USA\\
$^{2}$Department of Astronomy, Theoretical Astrophysics Center, University of California Berkeley, Berkeley, CA 94720, USA\\
$^{3}$Sydney Institute for Astronomy, School of Physics A28, University of Sydney, NSW 2006, Australia\\
$^{4}$ARC Centre of Excellence in All Sky Astrophysics in Three Dimensions (ASTRO-3D), Australia\\
$^{5}$Department of Astronomy, The University of Texas at Austin, 2515 Speedway, Stop C1400, Austin, TX 78712-1205, USA \\
$^{6}${Department of Physics and Astronomy, University of Alabama, Box 870324, Tuscaloosa, AL 35487-0324, USA}\\
$^{7}${Department of Physics and Astronomy and CIERA, Northwestern University, 2145 Sheridan Road, Evanston, IL 60208, USA}\\
}
\date{Accepted XXX. Received YYY; in original form ZZZ}
\pubyear{2019}

\begin{document}
\label{firstpage}
\pagerange{\pageref{firstpage}--\pageref{lastpage}}
\maketitle

\begin{abstract}
Surveys of the Milky Way (MW) and M31 enable detailed studies of stellar populations across ages and metallicities, with the goal of reconstructing formation histories across cosmic time. These surveys motivate key questions for galactic archaeology in a cosmological context: when did the main progenitor of a MW/M31-mass galaxy form, and what were the galactic building blocks that formed it? We investigate the formation times and progenitor galaxies of MW/M31-mass galaxies using the FIRE-2 cosmological simulations, including 6 isolated MW/M31-mass galaxies and 6 galaxies in Local Group (LG)-like pairs at $z=0$. We examine main progenitor ``formation'' based on two metrics: (1) transition from primarily \textit{ex-situ} to \textit{in-situ} stellar mass growth and (2) mass dominance compared to other progenitors. We find that the main progenitor of a MW/M31-mass galaxy emerged typically at $z\!\sim\!3\!-\!4$ ($11.6\!-\!12.2\Gyr$ ago), while stars in the bulge region (inner 2 kpc) at $z\!=\!0$ formed primarily in a single main progenitor at $z\!\lesssim\!5$ ($\lesssim\!12.6\Gyr$ ago). Compared with isolated hosts, the main progenitors of LG-like paired hosts emerged significantly earlier ($\Delta z\!\sim\!2$, $\Delta t\!\sim\!1.6\Gyr$), with $\sim\!4\times$ higher stellar mass at all $z\!\gtrsim\!4$ ($\gtrsim\!12.2\Gyr$ ago). This highlights the importance of environment in MW/M31-mass galaxy formation, especially at early times. On average, about 100 galaxies with $\Mstar\!\gtrsim\!10^5\Msun$ went into building a typical MW/M31-mass system. Thus, surviving satellites represent a highly incomplete census (by $\sim\!5\times$) of the progenitor population.
\end{abstract}


\graphicspath{{./}{Figures/}}

\section{Introduction}
\label{sec:intro}

The properties of stellar populations within a galaxy, including their age, elemental abundances, and kinematics, all provide rich insight into the galaxy's formation history.
Early studies of the kinematics of stars with highly radial orbits in the Milky Way (MW) suggested that these stars must have formed differently from those on more disk-like orbits, implying that the MW formed via a gravitational collapse \citep{Eggen62}.
Other early studies proposed that stars and clusters in the outer halo formed from material in proto-galaxies that continued to fall into the galaxy after the central regions already collapsed \citep[e.g.][]{Ostriker75, Searle78}.
We know today that the processes involved in galaxy formation are more elaborate \citep{Freeman02}.

Initial theories for galaxy formation invoked the dissipational collapse of gas in dark-matter (DM) halos \citep[e.g.][]{Rees77, White78, Fall80, Mo98}.
More recent works have examined the effects of galaxy mergers as well \citep[e.g.][]{Springel05, Robertson06, Stewart08, GarrisonKimmel18}.
In gas-rich mergers, stellar feedback plays an important role in retaining gas content prior to the merger as it can heat the interstellar medium (ISM) and redistribute gas throughout the galaxy, even to larger radii where the effects of gravitational torquing are not as strong \citep{Hopkins09}.
Without feedback, the gas easily is torqued and falls to the center of the gravitational well, where it gets consumed in a starburst.
This implies that some part of galactic disks must survive a merger process, and the thick disk of the MW likely survived a significant merging event, which could have deposited fresh gas into the MW and dynamically heated stars from the thick disk into the stellar halo \citep{Gallart19}.

The stellar halo of the MW is perhaps the best place to probe the remnants of its early formation process.
Various works have studied hierarchical formation of the stellar halo \citep[e.g.][]{Bullock01, Bullock05, Helmi08, Johnston08, Cooper10, Deason16}, showing that it occurs via the tidal disruption and accretion of many satellite dwarf galaxies.
For instance, using cosmological zoom-in simulations, \citet{Deason16} find that typically $1 - 2$ satellite galaxies contribute most of the accreted stellar material to a stellar halo.
More generally, they find that the majority of accreted metal-poor stars come primarily from `classical' dwarf galaxies ($\sim 40 - 80$ per cent) as opposed to `ultra-faint' dwarf galaxies (only $\sim 2 - 5$ per cent).
They also find a relation between the galaxy's progenitor mass and its satellite population at $z = 0$: galaxies with less massive progenitors tend to have more quiescent histories, as well as a less massive surviving satellite population, when compared to the more massive galaxies.
Similarly, examining the AURIGA simulations, \citet{Monachesi19} find a correlation with the number of `significant progenitors' (number of progenitors that contribute 90 per cent of the stellar halo mass) and the accreted mass in the halo, with more massive halos accreting smaller numbers of significant progenitors.
Studies of kinematically coherent structures in the MW's halo, like the Sagittarius stream \citep{Newberg2002, Majewski2003} and Gaia-Enceladus \citep{Helmi18, Belokurov18} clearly confirm this hierarchical formation scenario.
In particular, Gaia-Enceladus is thought to be comparable in stellar mass to the SMC ($\sim 6 \times 10^8 \Msun$), contributing most of the stars in the (inner) stellar halo \citep{Helmi18}.

Studies of old and/or metal-poor stars provide the best window into the early formation of the MW \citep[e.g.][]{Brook07, Scannapieco06, Deason16, Griffen18, Sestito19, Chiaki19}.
Current spectroscopic surveys (e.g. RAVE, GALAH, APOGEE, LAMOST) now provide elemental abundances and ages for stars across the MW \citep[e.g.][]{RAVE, GALAH, APOGEE, LAMOST}.
These surveys achieve high spectral resolution (up to $R \sim 30,000$) and signal-to-noise (up to $S/N > 100$), and are capable of observing stars with $\rm [Fe/H] < -2$.
Recently, the \textit{Pristine} survey has observed significant populations of stars at $\rm [Fe/H] < -3$, with promise of reaching to $\rm [Fe/H] < -4$ \citep{PRISTINE}.
Interestingly, using observations of metal-poor stars, \citet{Sestito19} found that a significant fraction of these stars with $\rm [Fe/H] < -4$ are on disk-like orbits in the MW.
Similar work using metal-poor stars from LAMOST also has proven useful in finding halo structures \citep[e.g.][]{Yuan19}.
A key question is where these metal-poor stars come from.
Did they form within the MW, or did they form in other dwarf galaxies that subsequently merged in?
If the latter, it would not make sense to say that they formed in the MW, or at least, its main progenitor.

Recently, using the FIRE-2 cosmological zoom-in baryonic simulations, \citet{ElBadry18} predicted that the oldest ($z_{\rm form} > 5$), metal-poor stars ([Fe/H] $\lesssim -2$) in the MW should be less centrally concentrated than stars that formed later, because (1) early merger events deposited stars that formed in dwarf galaxies on dispersion-supported orbits, and (2) stars that formed within the primary galaxy were heated to larger orbits via feedback-driven time-varying galactic potential.
A similar study using the APOSTLE simulations of MW/M31-like pairs by \citet{Starkenburg17} found comparable results for stars with $z_{\rm form} > 6.9$ and [Fe/H] $\lesssim -2.5$.

While current MW surveys give us detailed information about a star's position, kinematics, and elemental abundances, obtaining precise ages for stars remains challenging.
Current methods include fitting isochrones to stellar populations in colour-magnitude diagrams, studying oscillation modes of individual stars (astroseismology), using the rotation-age relation to infer ages (gyrochronology), and using detailed elemental abundances \citep[e.g.][]{Chaplin14, Martig16, Creevey17, SilvaAguirre18}.
Uncertainties in the ages of stars using the latter method can be as large as 40 per cent, but this improves if one can use a combination of methods \citep[e.g.][]{PLATO17}.
Gaia's second data release \citep[DR2;][]{Gaia18} now provides distance measurements for over 1.3 billion stars in the MW, allowing astronomers to measure isochrone ages.
For example, \citet{Gallart19}
suggest that the MW halo formed 50 per cent of its stars by $z \sim 4.2$ (12.3 Gyr ago) and the thick disk formed 50 per cent by $z \sim 2$ (10.5 Gyr ago).
Other analyses reported formation lookback times of the halo of $8 - 13 \Gyr$ \citep[e.g.][]{Schuster12, Hawkins14}, including different formation times for different halo populations \citep[e.g.][]{Ge16}.
Measurements of stellar ages in the MW bulge suggest that the stellar population is predominantly older than $\sim 10$ Gyr \citep[][and references therein]{Barbuy18}.

Cosmological galaxy simulations provide the best theoretical laboratories for understanding the full evolutionary histories of galaxies across cosmic time.
It remains unclear if the formation histories of MW/M31-mass galaxies depend on whether they are in isolated environments (with no other nearby companions of similar mass) or in LG-like environments (with a pair of massive galaxies at $\lesssim 1$ Mpc separation).
Most simulations of MW/M31-mass galaxies, either idealized or cosmological focus on a \textit{single isolated} galaxy/halo (e.g. AURIGA, NIHAO, ERIS, Caterpillar), not in a Local Group (LG)-like MW+M31 pair \citep{AURIGA, NIHAO, ERIS, Caterpillar}.
Exceptions include the ELVIS DM-only (DMO) simulation suite, which include 24 MW/M31-mass halos in LG-like pairs and a mass-matched sample of 24 isolated halos \citep{GarrisonKimmel14}, the cosmological baryonic simulations of LG-like pairs from the APOSTLE suite \citep{APOSTLE}, and the simulated LG analogues from the CLUES project \citep{CLUES}.
Furthermore, semi-analytic models applied in a cosmological settings provide complementary tools to undersand MW-mass formation histories and their environmental dependence.
For instance, the semi-analytical code GAlaxy MErger Tree and Evolution (GAMETE, and the similar code GAMESH which accounts for radiative transfer) allows for a detailed modeling for things such as the gas evolution within a MW-mass galaxy, the evolution of the galactic SFR, and the formation/evolution of early Population III (Pop III) stars \citep{Salvadori07, deBennassuti14, Graziani15}.

In addition to potential differences in host galaxy properties in LG-like versus isolated environments, it is also imperative to understand potential differences in their satellite populations.
While observations of the LG have driven most of our knowledge of dwarf galaxy populations, recent observational campaigns aim to measure satellite populations around (primarily more isolated) MW-mass galaxies such as M94 \citep{Smercina18}, M101 \citep{Danieli17}, M81 \citep{Karachentsev14}, and Centaurus A \citep{Muller19}.
The Satellites Around Galactic Analogs (SAGA) survey\footnote{See the SAGA survey web site: http://sagasurvey.org} also is observing satellite populations around (mostly) isolated MW-mass galaxies down to the luminosity of Leo I ($\rm M_r < -12.3$; $\Mstar \approx 5 \times 10^6 \Msun$), with a predicted sample size of 100 galaxies \citep{Geha17}.
Other campaigns instead focus on groups of galaxies out to $\sim 40$ Mpc \citep[e.g.][]{Kourkchi17}.
In connecting these observations with those of the LG, we must understand: \textit{does environment play a role in the satellite galaxy populations, and thus building blocks, of MW/M31-mass galaxies across cosmic time?}

In this work, we use FIRE-2 simulations of 12 MW/M31-mass galaxies to investigate their cosmological hierarchical formation histories.
We quantify these galaxies' building blocks across cosmic time and determine when their main progenitors formed/emerged.
We investigate the times at which (1) the \textit{stellar} mass growth of the most massive progenitor transitions from being dominated by ex-situ stars (via mergers) to in-situ star formation, and (2) when the most massive galaxy starts to dominate in stellar mass compared to other progenitor galaxies.
When a progenitor galaxy satisfies either of these two criteria, we define it to be the `main' progenitor, which is distinct from the other progenitor galaxies that continue to merge in.
There are many definitions of host galaxy `formation', such as when the main galaxy reaches a fraction of its mass at $z = 0$; here we refer to formation specifically as when a single main progenitor galaxy emerges in its environment, based both on its in-situ star formation (Section~\ref{sec:insitu}) and how its stellar mass compares to its neighbors (Section~\ref{sec:mrs}).
We also emphasize that in our analysis, we focus on the formation of the host galaxy generally, and not specifically on the formation of a given component, such as the thin or thick disk.

Our work, however, is not the first to investigate general properties of the progenitors of MW-mass galaxies and their satellites.
Many studies, including those from \citet{Dixon18, Safarzadeh18, Graziani17, Magg18, deBennassuti17}, have focused on the high-$z$ Universe to study star formation near the epoch of reionization, the metallicity distribution function, identifying ultra-faint dwarf galaxy progenitors, and Pop III stars to understand the early MW environment.

The main questions that we address are:
\renewcommand{\labelenumi}{\alph{enumi})}
\begin{enumerate}
    \item What were the building blocks (progenitor galaxies) of MW/M31-mass galaxies, and how many were there across cosmic time?
    \item When did the main progenitor of a MW/M31-mass galaxy form/emerge?
    \item Does the formation of MW/M31-mass galaxies depend on their environment, specifically, comparing isolated hosts to those in LG-like pairs?
\end{enumerate}

\section{Methods}
\label{sec:methods}

\begin{table*}
	\centering
	\begin{threeparttable}
	\caption{
	Properties of the 12 host galaxies in the FIRE-2 simulation suite that we analyze. Column list: name; stellar mass (M$_{\rm{star,90}}$) within R$_{\rm star,90}$; disk radius enclosing 90 per cent of the stellar mass within 20 kpc (R$_{\rm star,90}$); halo virial mass ($\Mthm$); halo virial radius ($\Rthm$); redshift when the galaxy reached 50 per cent ($z_{0.5}$) or 10 per cent ($z_{0.1}$) of its stellar mass at $z = 0$; redshift when the cumulative fraction of stars that formed in-situ exceeded 0.5 when selecting stars at $z = 0$ within host-centric distances of 15 kpc ($z_{\rm in-situ, 15}$) and 2 kpc ($z_{\rm in-situ, 2}$; redshift when which the most massive progenitor exceeded a 3:1 stellar mass ratio with respect to the second most massive progenitor ($z_{\rm MR, 15}$); and the paper that introduced each simulation. Hosts with names starting with `m12' are isolated hosts from the Latte suite, while the rest are in Local Group (LG)-like pairs from the ELVIS on FIRE suite.
	}
	\begin{tabular}{|c|c|c|c|c|c|c|c|c|c|c|}
		\hline
		\hline
		Name & $\rm M_{\rm star,90}$ & R$_{\rm star,90}$ & $\Mthm$ & $\Rthm$ & $z_{0.5}$ & $z_{0.1}$ & $z_{\rm in-situ, 15}$ & $z_{\rm in-situ, 2}$ & $z_{\rm MR, 15}$ & Ref\\
		 & [$10^{10} \Msun$] & [$\kpc$] & [$10^{12} \Msun$] & [$\kpc$] & & & & & &  \\
		\hline
        m12m & 10.0 & 11.6 & 1.6 & 371 & 0.6 & 1.5 & 1.9 & 2.1 & 1.5 & A \\
        Romulus & 8.0 & 12.9 & 2.1 & 406 & 0.7 & 1.4 & 5.3 & > 6.0 & 1.7 & B \\
        m12b & 7.3 & 9.0 & 1.4 & 358 & 0.6 & 1.7 & 2.4 & 4.2 & 3.7 & C \\
        m12f & 6.9 & 11.8 & 1.7 & 380 & 0.5 & 2.0 & 3.7 & > 6.0 & 2.9 & D \\
        Thelma & 6.3 & 11.2 & 1.4 & 358 & 0.4 & 1.1 & 1.7 & 1.7 & 4.4 & C \\
        Romeo & 5.9 & 12.4 & 1.3 & 341 & 1.0 & 2.6 & > 6.0 & > 6.0 & > 6.0 & C \\
        m12i & 5.5 & 8.5 & 1.2 & 336 & 0.6 & 1.7 & 3.1 & 3.5 & 1.7 & E \\
        m12c & 5.1 & 9.1 & 1.4 & 351 & 0.5 & 1.3 & 1.9 & 5.3 & 3.1 & C \\
        m12w & 4.8 & 7.3 & 1.1 & 319 & 0.4 & 1.4 & 2.7 & 3.4 & 1.2 & F \\
        Remus & 4.0 & 11.0 & 1.2 & 339 & 0.9 & 2.7 & 3.7 & > 6.0 & 1.2 & B \\
        Juliet & 3.3 & 8.1 & 1.1 & 321 & 0.8 & 2.4 & 5.3 & 5.5 & 4.7 & C \\
        Louise & 2.3 & 11.2 & 1.2 & 333 & 0.9 & 2.5 & 4.9 & 5.0 & 3.5 & C \\
		\hline
	\end{tabular}
\label{tab:hosts}
\begin{tablenotes}
\item \textit{Note:} The references for each host are: A: \citet{Hopkins18}, B: \citet{GarrisonKimmel19b}, C: \citet{GarrisonKimmel19a}, D: \citet{GarrisonKimmel17}, E: \citet{Wetzel16}, and F: \citet{Samuel19}.
\end{tablenotes}
\end{threeparttable}
\end{table*}

\subsection{FIRE-2 Simulations of Milky Way- and M31-mass Galaxies}
\label{sec:sims}

We use cosmological zoom-in baryonic simulations of MW/M31-mass galaxies from the Feedback In Realistic Environments (FIRE) project\footnote{See the FIRE project web site: http://fire.northwestern.edu} \citep{Hopkins18}.
We ran these simulations using the \texttt{Gizmo} $N$-body gravitational plus hydrodynamics code \citep{Hopkins15}, with the mesh-free finite-mass (MFM) hydrodynamics method and the FIRE-2 physics model \citep{Hopkins18}.
FIRE-2 includes several radiative cooling and heating processes for gas such as free-free emission, photoionization/recombination, Compton scattering, photoelectric, metal-line, molecular, fine-structure, dust-collisional, and cosmic-ray heating across a temperature range of $10 - 10^{10} \K$.
This includes the spatially uniform, redshift-dependent cosmic UV background from \cite{FaucherGiguere09}, for which HI reionization occurs at $z_{\rm reion} \sim 10$.
The simulations self-consistently generate and track 11 elemental abundances (H, He, C, N, O, Ne, Mg, Si, S, Ca, Fe), including sub-grid diffusion of these abundances in gas via turbulence \citep{Hopkins16, Su17, Escala18}.

Stars form from gas that is self-gravitating, Jeans unstable, molecular \citep[following][]{Krumholz11}, and dense ($n_H$ > 1000 cm$^{-3}$).
Once a star particle forms, inheriting mass and elemental abundances from its progenitor gas element, it represents a single stellar population, assuming a \cite{Kroupa01} initial mass function, and it evolves along stellar population models from \texttt{STARBURST99 v7.0} \citep{Leitherer99}.
FIRE-2 simulations include several different feedback processes, including core-collapse and Ia supernovae, mass loss from stellar winds, and radiation, including radiation pressure, photoionization, and photo-electric heating.

We generated cosmological zoom-in initial conditions for each simulation at $z \approx 99$, embedded within periodic cosmological boxes of lengths $70.4 - 172$ Mpc using the code \texttt{MUSIC} \citep{Hahn11}.
We saved 600 snapshots down to $z = 0$, which are spaced every $\approx 25$ Myr.
All simulations assume flat $\Lambda$CDM cosmology with parameters consistent with \citet{Planck18}.
Specifically, the Latte suite (excluding m12w) used $\Omega_{\rm m} = 0.272$, $\Omega_{\rm b} = 0.0455$, $\sigma_{\rm 8} = 0.807$, $n_{\rm s} = 0.961$, $h = 0.702$.
Thelma \& Louise and Romulus \& Remus both used the same cosmology as in the original ELVIS DMO suite: $\Omega_{\rm m} = 0.266$, $\Omega_{\rm b} = 0.0449$, $\sigma_{\rm 8} = 0.801$, $n_{\rm s} = 0.963$, $h = 0.71$.
Finally, Romeo \& Juliet and m12w both used $\Omega_{\rm m} = 0.31$, $\Omega_{\rm b} = 0.048$, $\sigma_{\rm 8} = 0.82$, $n_{\rm s} = 0.97$, $h = 0.68$. 

In this work, we analyze 12 MW/M31-mass galaxies; Table~\ref{tab:hosts} lists their properties.
6 galaxies are from the Latte suite of isolated MW/M31-mass galaxies, introduced in \cite{Wetzel16}.
We selected these halos with the following two criteria: (1) $\Mthm = 1-2 \times 10^{12} \Msun$ \citep[e.g.][]{Bland-Hawthorn16}, and (2) no similar-mass halo within $5 \times \Rthm$ (for computational efficiency).
Here, $\Mthm$ refers to the total mass within a radius, $\Rthm$, containing 200 times the mean matter density of the Universe.
We chose galaxy m12w with one additional criterion: having an LMC-mass satellite at $z \sim 0$ in the pilot DM-only simulation \citep[see][]{Samuel19}.
The Latte simulations have gas and \textit{initial} star particle masses of $7100 \Msun$, although because of stellar mass loss, the typical star particle at $z = 0$ is $\sim 5000 \Msun$.
DM particles have masses of $3.5 \times 10^4 \Msun$.
Gas elements use fully adaptive force softening, equal to their hydrodynamic smoothing, that adapts down to 1 pc.
The gravitational softening lengths for star and DM particle are fixed at 4 and 40 pc (Plummer equivalent), comoving at $z > 9$ and physical thereafter.

We also use 6 galaxies from the ``ELVIS on FIRE'' suite of 3 Local Group (LG)-like MW+M31 pairs \citep{GarrisonKimmel19b, GarrisonKimmel19a}, which were selected with the following criteria: (1) two neighboring halos each with mass $\Mthm = 1 - 3 \times 10^{12} \Msun$, (2) total LG mass of $2 - 5 \times 10^{12} \Msun$, (3) center separation of $600 - 1,000 \kpc$ at $z = 0$, (4) radial velocities of $v_{\rm rad} < 0 \kmsi$ at $z = 0$, and (5) no other massive halos within 2.8 Mpc of either host center.
The ELVIS on FIRE simulations have $\approx 2 \times$ better mass resolution than Latte, with initial star/gas particle masses of $3,500 - 4,000 \Msun$.

\citet{Hopkins18} examined the effect of mass resolution on the formation histories of both m12i and m12m, finding differences $\lesssim 10-20$ per cent in their star formation histories (SFHs) comparing star particle masses of $7,000 \Msun$ to $56,000 \Msun$.
We thus expect that the factor of $\approx 2 \times$ resolution difference between the Latte and ELVIS suites should not significantly affect their SFHs.

We simulated these galaxies using the cosmological zoom-in technique \citep{Onorbe14}: see \citet{Wetzel16}, \citet{Hopkins18}, and \citet{GarrisonKimmel19b} for more detail.
We emphasize that we selected these halos solely using the parameters above, with no prior on their formation/merger histories or satellite populations (other than m12w).
Furthermore, the numerical implementation that we used to generate these simulations, the \texttt{Gizmo} source code and FIRE-2 physics model, were the same across all simulations.
Thus, these 12 hosts should reflect random/typical samplings of MW/M31-mass formation histories within their mass and environmental selection criteria.
However, as we show below, LG-like versus isolated host selection does lead to systematically different formation histories.

Both the Latte and ELVIS on FIRE simulation suites form MW/M31-mass galaxies with realistic populations of satellite galaxies, in terms of their stellar masses and velocity dispersions (dynamical masses) \citep{Wetzel16, GarrisonKimmel19b}, radial distributions \citep{Samuel19}, and star-formation histories \citep{GarrisonKimmel19a}.
The MW/M31-mass host galaxies in the simulations also show a range of morphologies \citep{GarrisonKimmel18, ElBadry18_gas}, with properties that broadly agree with the MW and M31, such as the stellar-to-halo mass relation \citep{Hopkins18}, disk structure \& gas mass \citep{Sanderson18}, age \& metallicity gradients \citep{Ma17}, and stellar halos \citep{Bonaca2017, Sanderson2018}.
The papers above provide detailed properties for at least some of the hosts.
For example, \citet{Ma17} investigated age and metallicity gradients, as well as the disk structure, of m12i.
\citet{Sanderson2018} presented surface densities of both the disk and bulge, and halo stellar masses for \textit{all} simulations used in this work.
\citet{Sanderson18} lists the gas masses, disk scale heights, and SFR (at $z = 0$) for m12f, m12i, and m12m.
Several upcoming works will present gas and stellar metallicities (Bellardini et al., in prep), and the evolution of SFR across redshift (Yu et al., in prep) for all of these simulated hosts.

\subsection{Identifying Halos and Galaxies}
\label{sec:rockstar}

We use the \texttt{ROCKSTAR} 6-D halo finder \citep{Behroozi13a} to identify DM (sub)halos, using $\Mthm$ as our halo definition.
We generate a halo catalog using only DM particles at each of the 600 snapshots and use \texttt{CONSISTENT-TREES} \citep{Behroozi13b} to construct merger trees.
Given the conservatively large zoom-in volume that we generate for each host, all of the halos that we examine have zero contamination by low-resolution DM particles.

We briefly summarize the method that we use to assign star particles to halos in post-processing; see \cite{Necib18} and \cite{Samuel19} for details.
Given each (sub)halo's radius and maximum circular velocity, $\Rhalo$ and $\Vcircmax$, as returned by \texttt{ROCKSTAR}, we identify star particles whose positions lie within 0.8 $\Rhalo$ (out to a maximum distance of 30 kpc) and whose velocities are within 2 $\Vcircmax$ of the (sub)halo's center-of-mass velocity.
After this, we keep star particles if they meet the following two criteria.
First, their positions are within 1.5 R$_{\rm star,90}$ (the radius that encloses 90 per cent of the stellar mass) of both the center-of-mass position of current member star particles and the halo center.
This criterion ensures that the galaxy's center of mass coincides with the halo's center of mass.
Second, their velocities are within 2 $\sigma_{\rm vel,star}$, the velocity dispersion of current member star particles, of the center-of-mass velocity of member star particles.
We iterate these two criteria until the stellar mass converges to within 1 per cent.
We then keep halos with at least 6 star particles, stellar density $> 300 \Msun \kpc^{-3}$ (at R$_{50}$, the radius which encloses 50 per cent of the stellar mass), and halo bound mass fraction > 40 per cent.
Henceforth, when we refer to a galaxy's stellar mass, we mean the mass that we calculate from this process.
We checked that none of our results change significantly if we use other ways of measuring stellar mass, such as the mass within R$_{\rm star,90}$.

Our software for reading and analyzing halo catalogs, including assigning star particles, is available via the \texttt{HaloAnalysis} package\footnote{https://bitbucket.org/awetzel/halo\_analysis}, and our software for reading and analyzing particles from Gizmo snapshots is available via the \texttt{GizmoAnalysis} package\footnote{https://bitbucket.org/awetzel/gizmo\_analysis}; we first developed and used these packages in \citet{Wetzel16}.

\subsection{Selecting Progenitor Galaxies}
\label{sec:select}

In our analysis, we impose two additional criteria to select galaxies of interest.
First, we examine only galaxies with stellar mass $\Mstar \geq 10^5 \Msun$, which corresponds to $\geq 14 - 29$ star particles, depending on simulation resolution.
Second, we analyze only galaxies that are progenitors of each MW/M31-mass system at $z = 0$, selecting stars at various host-centric distances at $z = 0$ to probe the formation histories of different regions of the host galaxy/halo.
Specifically, we select progenitor galaxies that contribute star particles to the following spherical distances, $d$, with respect to the center of each MW/M31-mass galaxy at $z = 0$:
\renewcommand{\labelenumii}{\roman{enumii}}
\begin{enumerate}
\item $d(z = 0) < 300 \kpc$ (hereon $d_{300}$), corresponding to the entire host halo system (virial region), including the entire host galaxy, stellar halo, and surviving satellite galaxies;
\item $d(z = 0) < 15 \kpc$ ($d_{15}$), corresponding to the entire host galaxy (bulge and disk) plus inner stellar halo;
\item $d(z = 0) < 2 \kpc$ ($d_2$), corresponding to an inner bulge region.
\end{enumerate}
While these represent relatively simple spherical distance selections, we also investigated a `disk' selection, by selecting stars at $z = 0$ within $R = 4-15 \kpc$ and $|Z| < 2 \kpc$ (in cylindrical coordinates), as well as requiring stars to be on co-rotating disk-like circular orbits via Toomre diagram selection.
While this reduces the overall amount of accreted stars (as expected), given our method to select progenitor galaxies (see next paragraph), this selection generally did not lead to significant differences in our results compared to $d_{15}$.

Having defined the star particles in each region at $z = 0$, to select progenitor galaxies at a given redshift, we compute how much stellar mass a galaxy at a given redshift contributed to these host-centric distances at $z = 0$, relative to the galaxy's total stellar mass at that redshift.
We define this as the `contribution fraction'.
To be a progenitor, we require that a galaxy has a contribution fraction greater than 1 per cent, that is, at least 1 per cent of its mass (at a given redshift) ends up within the host-centric region.
We checked how our metrics for progenitor formation (described in Sections~\ref{sec:insitu} and \ref{sec:mrs}, and summarized in \ref{sec:reds}) changed as we varied this contribution fraction requirement.
For $d_{300}$, our results are not sensitive to it, because all contribution fractions are near 100 per cent, indicating that galaxies that contribute to the host halo contribute essentially all of their stars.
For $d_2$, and to a lesser extent $d_{15}$, the contribution fractions are more broadly spread throughout 0 - 100 per cent, because stars from infalling galaxies get deposited across a range of $d$ by $z = 0$.
We thus use 1 per cent contribution fraction as a conservative minimum, though we note that increasing this minimum would decrease the number of progenitors to the inner galaxy.

\section{Results}
\label{sec:results}

\begin{figure*}
\centering
\begin{tabular}{c @{\hspace{0.2ex}} c @{\hspace{0.2ex}} c}
    \includegraphics[width=0.32\linewidth]{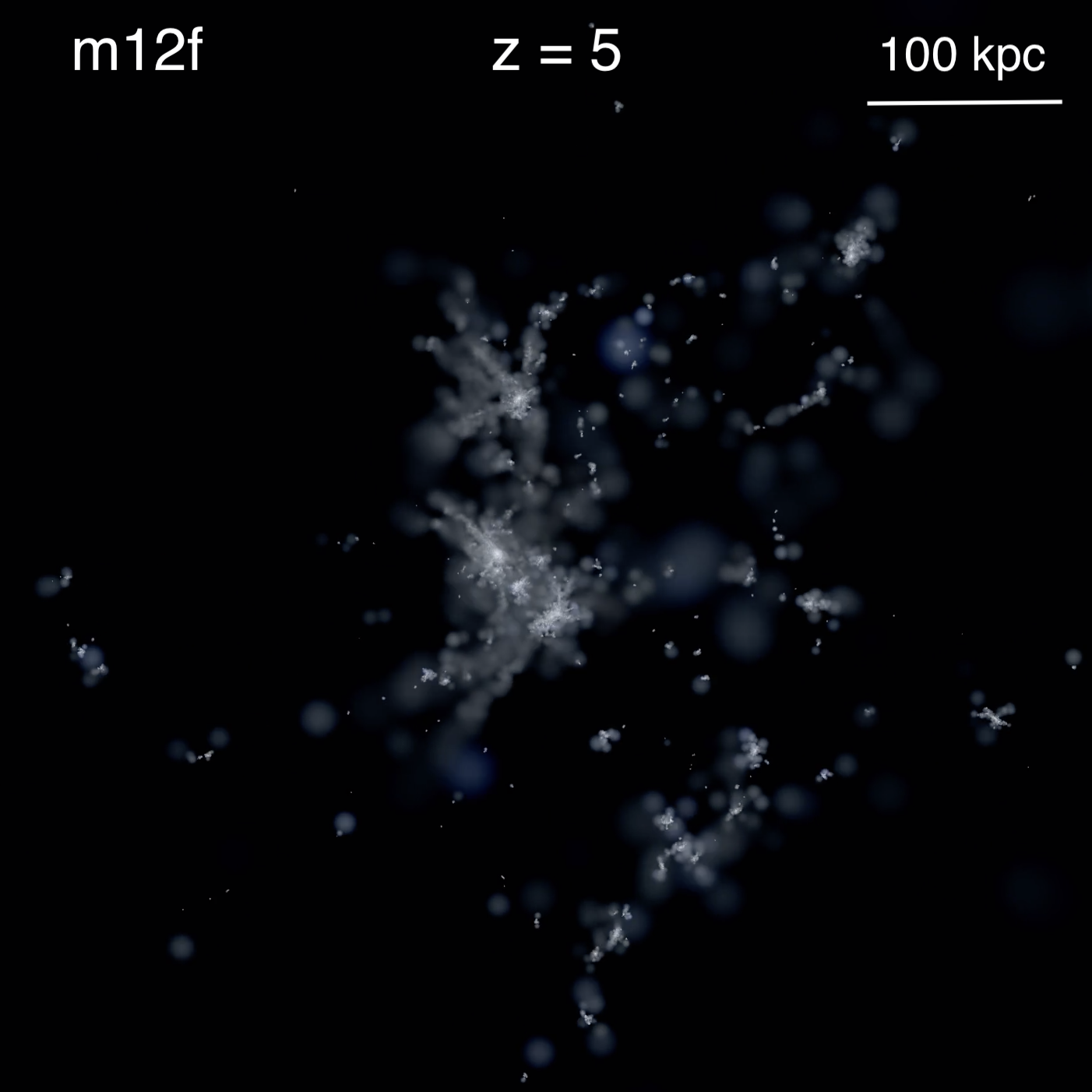}&
    \includegraphics[width=0.32\linewidth]{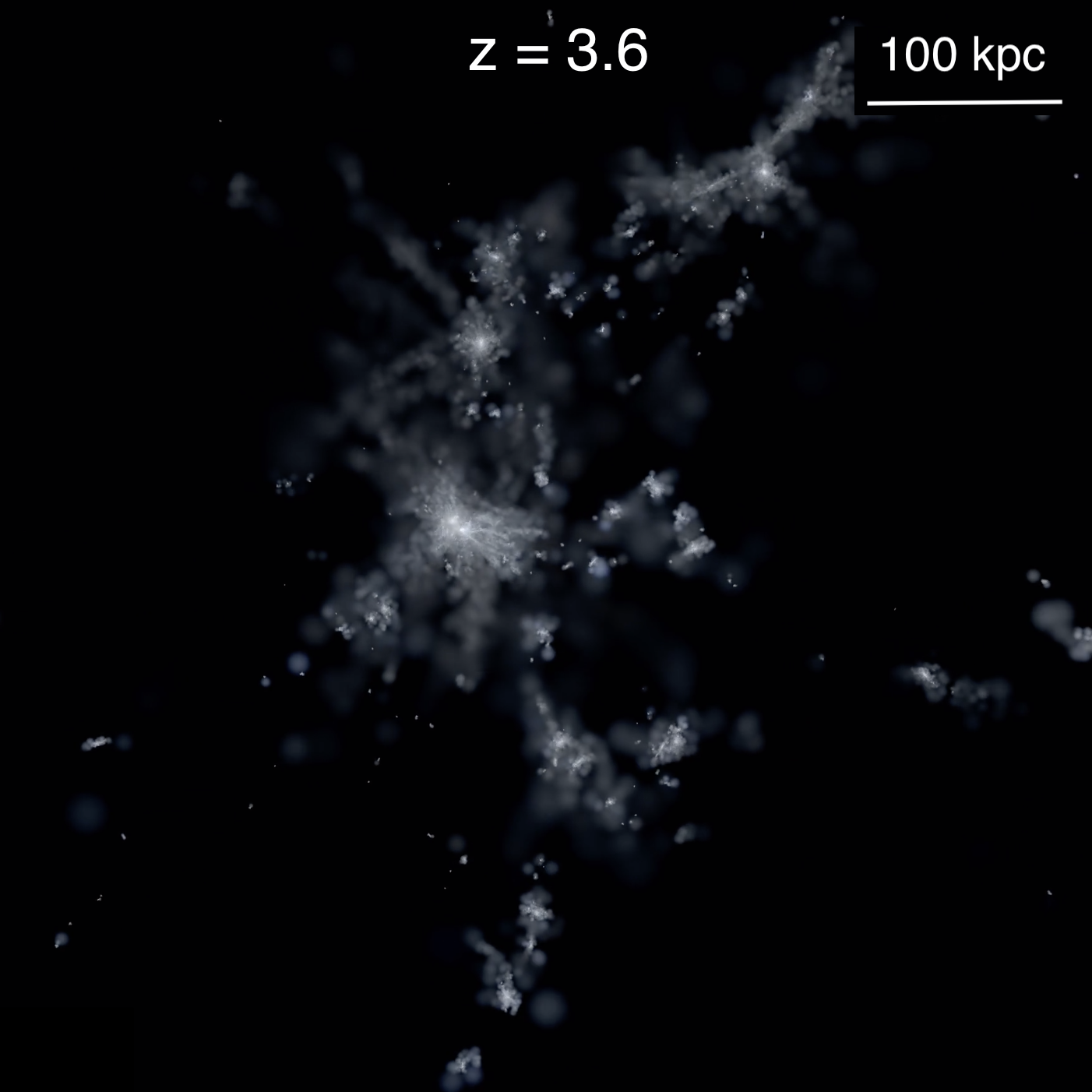}&
    \includegraphics[width=0.32\linewidth]{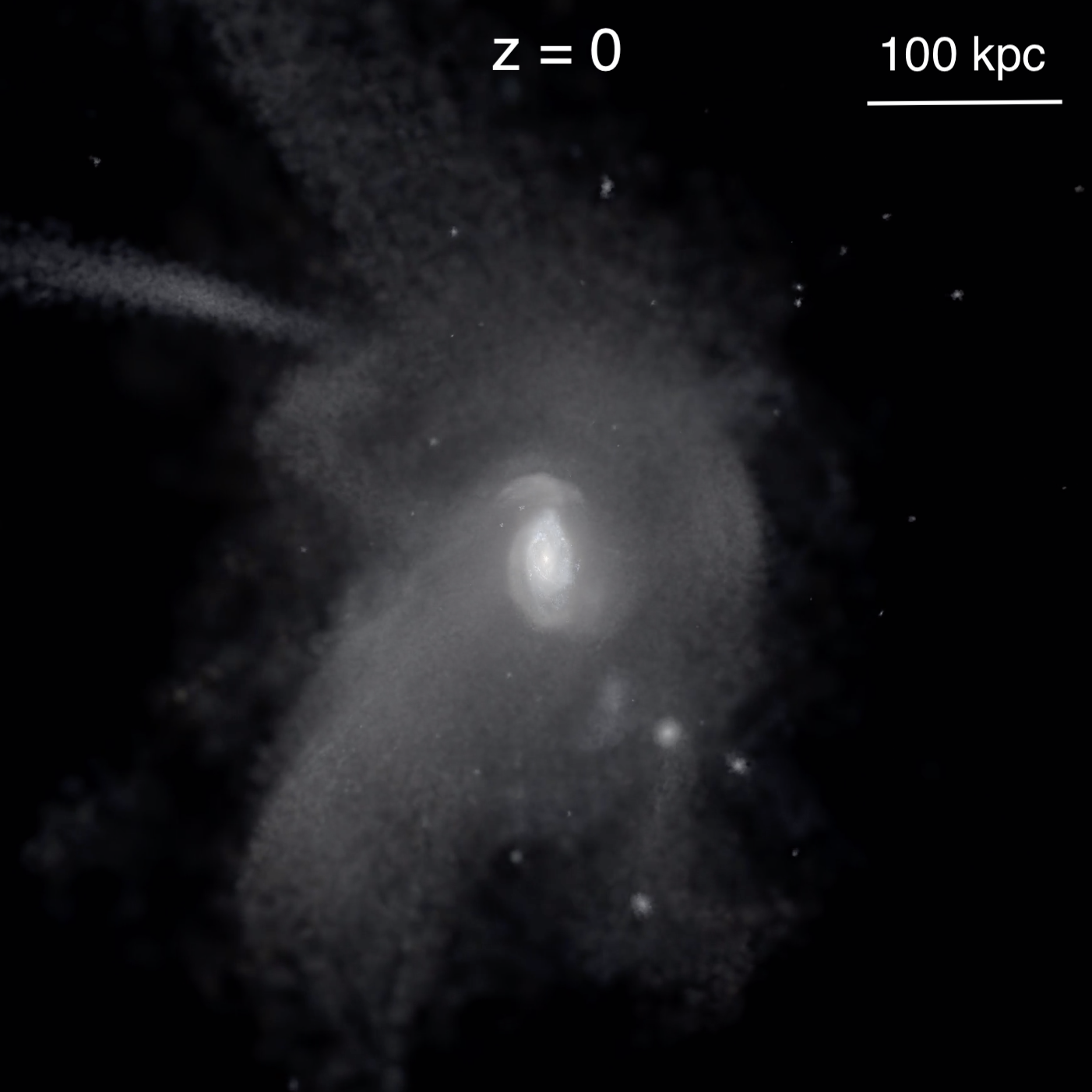} \\ \\
    \includegraphics[width=0.32\linewidth]{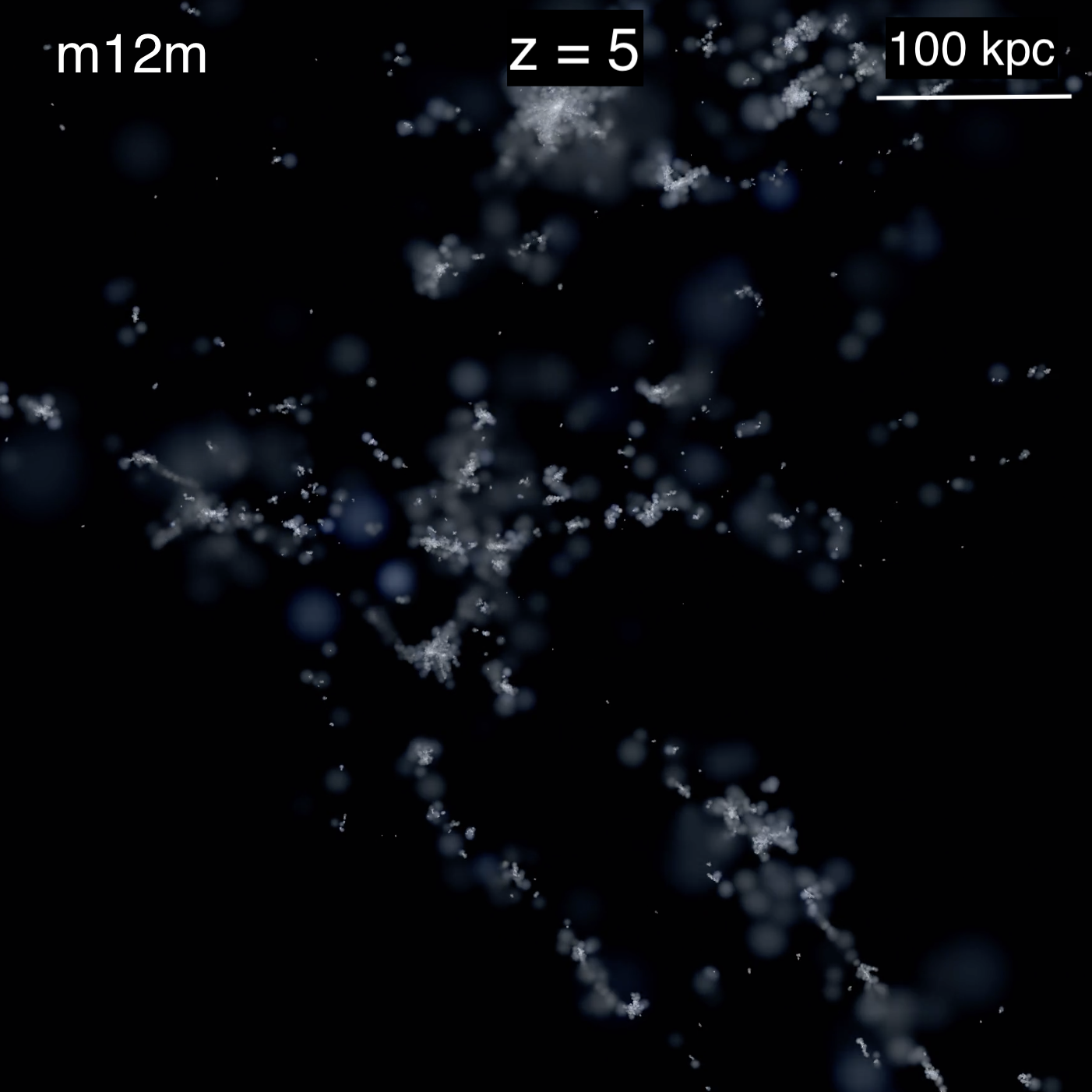}&
    \includegraphics[width=0.32\linewidth]{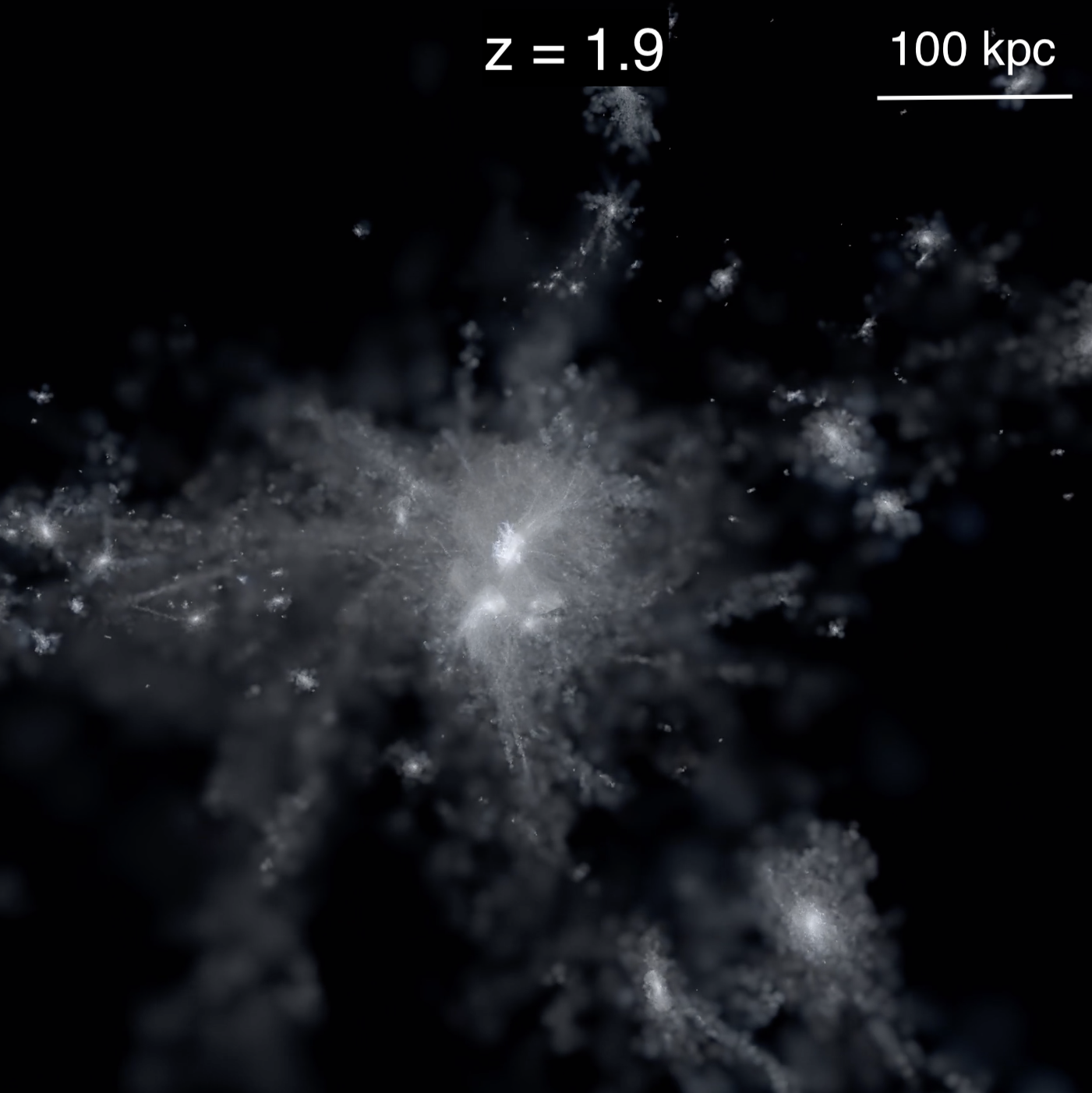}&
    \includegraphics[width=0.32\linewidth]{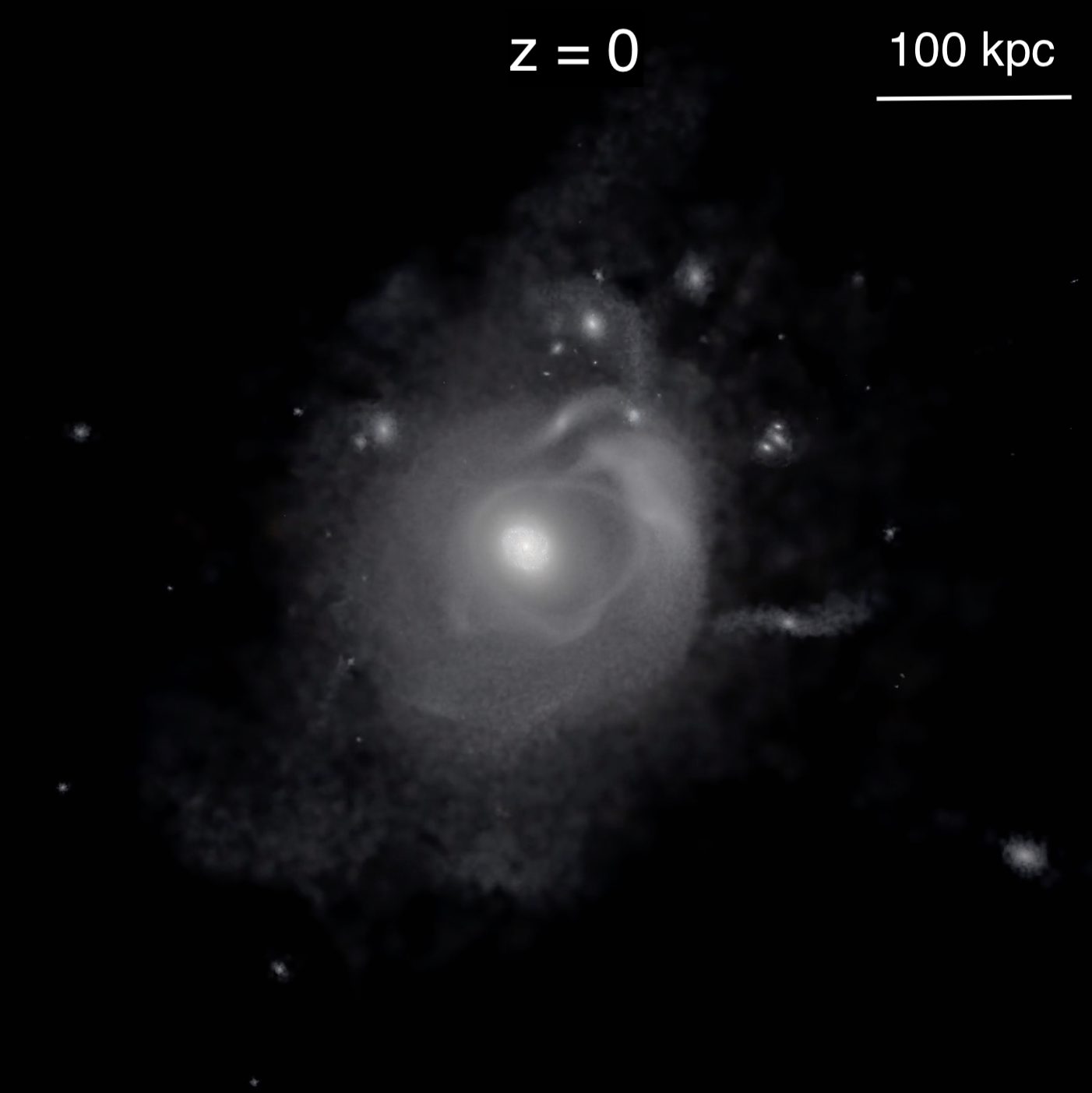}
\end{tabular}
\vspace{-2 mm}
\caption{
Synthetic Hubble Space Telescope u/g/r composite stellar images of all star particles (ignoring dust attenuation) out to a distance of 300 kpc physical (roughly $\Rthm$ at $z = 0$) around the most massive progenitor (MMP) at each redshift, for two simulations that bracket the main progenitor formation times across our isolated hosts.
Each star particle is represented as a spherical cloud, with size representative of the local density of star particles.
Intensity is logarithmic with stellar density, the visible regions ranging from $10^{-9} - 3 \times 10^{-2} \ \rm M_{\odot} \ pc^{-3}$.
Left to right, the panels show images at $z = 5$, $z = z_{\rm form}$ (when the main progenitor `formed' or `emerged'), and at $z = 0$.
We define main progenitor formation when the majority of stars for $d_{15}$ have formed in-situ in a single progenitor (see Section~\ref{sec:insitu}); this is similar to when the stellar mass ratio of the most massive to second most massive progenitor exceeds 3:1 (see Section~\ref{sec:mrs}).
The top row shows m12f, our earliest forming isolated host, and the bottom row shows m12m, our latest forming isolated host.
\textit{Prior to $z_{\rm form}$, there was little meaningful sense of a single `main' progenitor; instead, there was a collection of similar-mass progenitor galaxies.}
}
\label{fig:hist}
\end{figure*}

To provide visual context to our analysis, Fig.~\ref{fig:hist} shows synthetic Hubble Space Telescope u/g/r composite stellar images of two of our isolated hosts---m12f (top) and m12m (bottom)---using \texttt{STARBURST99} to determine the SED for each star particle (given its age and metallicity) and following the ray-tracing methods in \citet{Hopkins05, Hopkins18} for an observer 1 Mpc away.
These images do not include dust attenuation.
Each star particle is represented as a spherical cloud, with size representative of the local density of star particles.
These images show stellar luminosity along a given line of sight using a logarithmic color scale, with visible stellar densities ranging from $10^{-9} - 3 \times 10^{-2} \ \rm M_{\odot} \ pc^{-3}$.

Left to right, each panel shows different times throughout the formation of each MW/M31-mass galaxy, including all stars out to 300 kpc (physical) from the most massive progenitor.
At $z = 0$ (right), the MW/M31-mass host galaxy is clearly dominant, including 16-27 resolved surviving satellite galaxies and an extended stellar halo with stellar streams from disrupted satellites.
Some of the visible streams formed via outflows of gas \citep{Yu20}, while others form via satellite disruption.
However, at $z = 5$ (left), the system was composed of a collection of many similar-mass galaxies that eventually merge together.
Because many galaxies had similar mass and none of them alone dominated the stellar mass growth, there was no meaningful single `main' progenitor at this time.
The middle panel shows each host when its main progenitor `formed/emerged', which we define as when the cumulative stellar mass was dominated by in-situ formation in a single progenitor (see Section~\ref{sec:insitu}).
We define in-situ formation as being star formation which occurs in the most massive progenitor (MMP); we discuss details of how we calculate the \textit{in-situ fraction} in the beginning of Section~\ref{sec:insitu}.
Similarly, ex-situ formation is defined as that which occurs in any other progenitor galaxy; we discuss details about how we determine the \textit{ex-situ fraction} in Section~\ref{sec:mfs}.
Compared to the panels on the left, we see the emergence of a main progenitor galaxy that dominates its local environment in mass by more than 3:1 (see Section~\ref{sec:mrs}).

\subsection{What was the mass growth of the most massive progenitor and does it depend on environment?}
\label{sec:mmp growth}

First, we examine the stellar mass growth history of the MMP of each MW/M31-mass galaxy from $z = 0$ back to $z = 7$.
Following the criteria above, we identify the progenitor galaxy with the highest stellar mass at each snapshot (according to our halo catalog) and label it as the MMP.
This is different from using the halo merger tree to track the galaxy back in time, because DM halo and stellar masses grow at different rates, so the most massive halo is not always the most massive galaxy, especially at early times before a main progenitor has emerged.
Furthermore, as we argue in Section~\ref{sec:insitu}, it does not make sense to think about the MMP as the single `main' progenitor before the `formation' redshift of each host.
Our goal here is to set the stage by providing the relevant mass scale of the MMP across cosmic time while highlighting environmental differences.
Also, the results in this subsection do not depend on the details of host-centric distance selection at $z = 0$, because the MMP contributes significant mass inside all of our distance cuts.

\begin{figure*}
\centering
\begin{tabular}{c @{\hspace{-0.1ex}} c}
\includegraphics[width=0.49\linewidth]{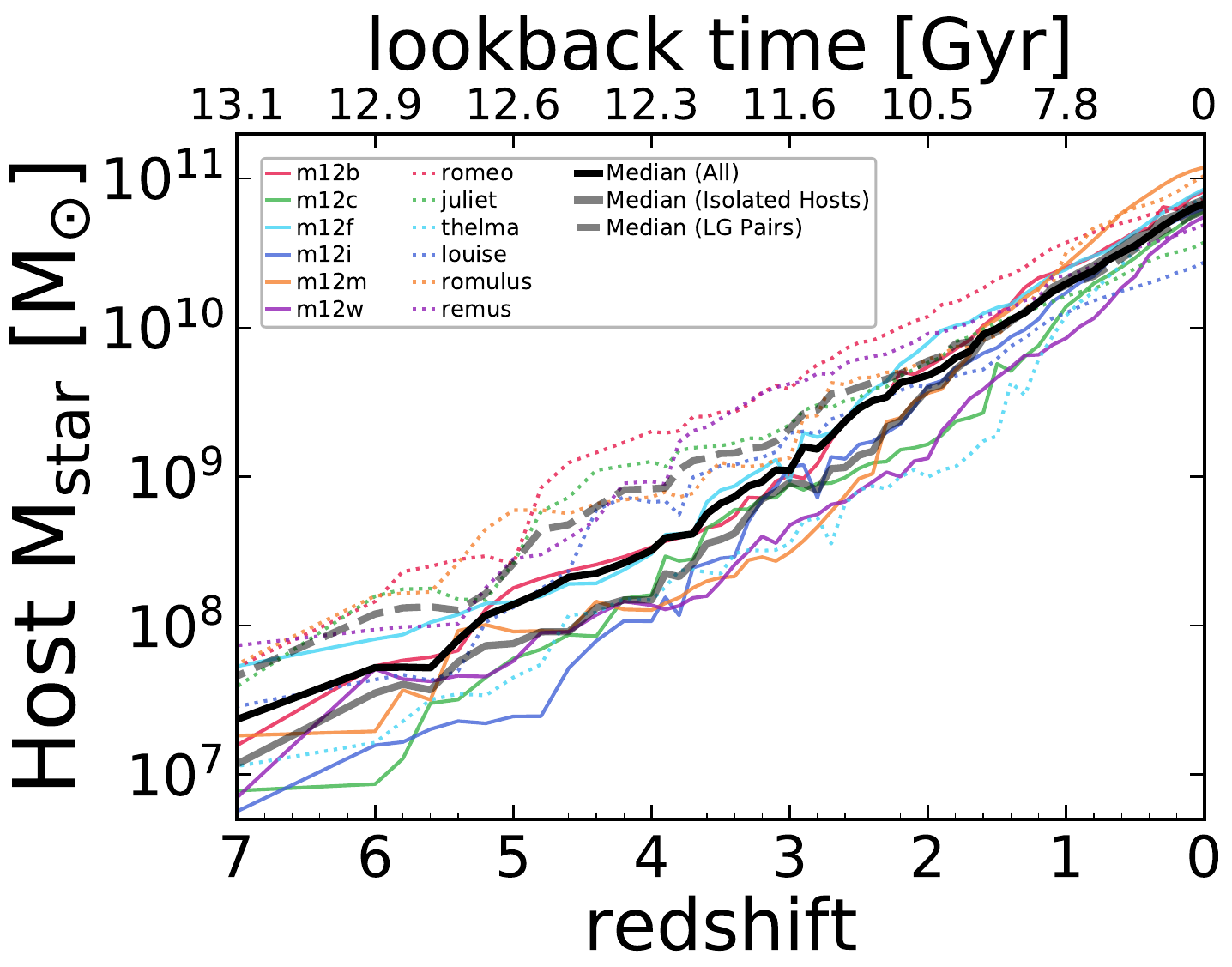} &
\includegraphics[width=0.49\linewidth]{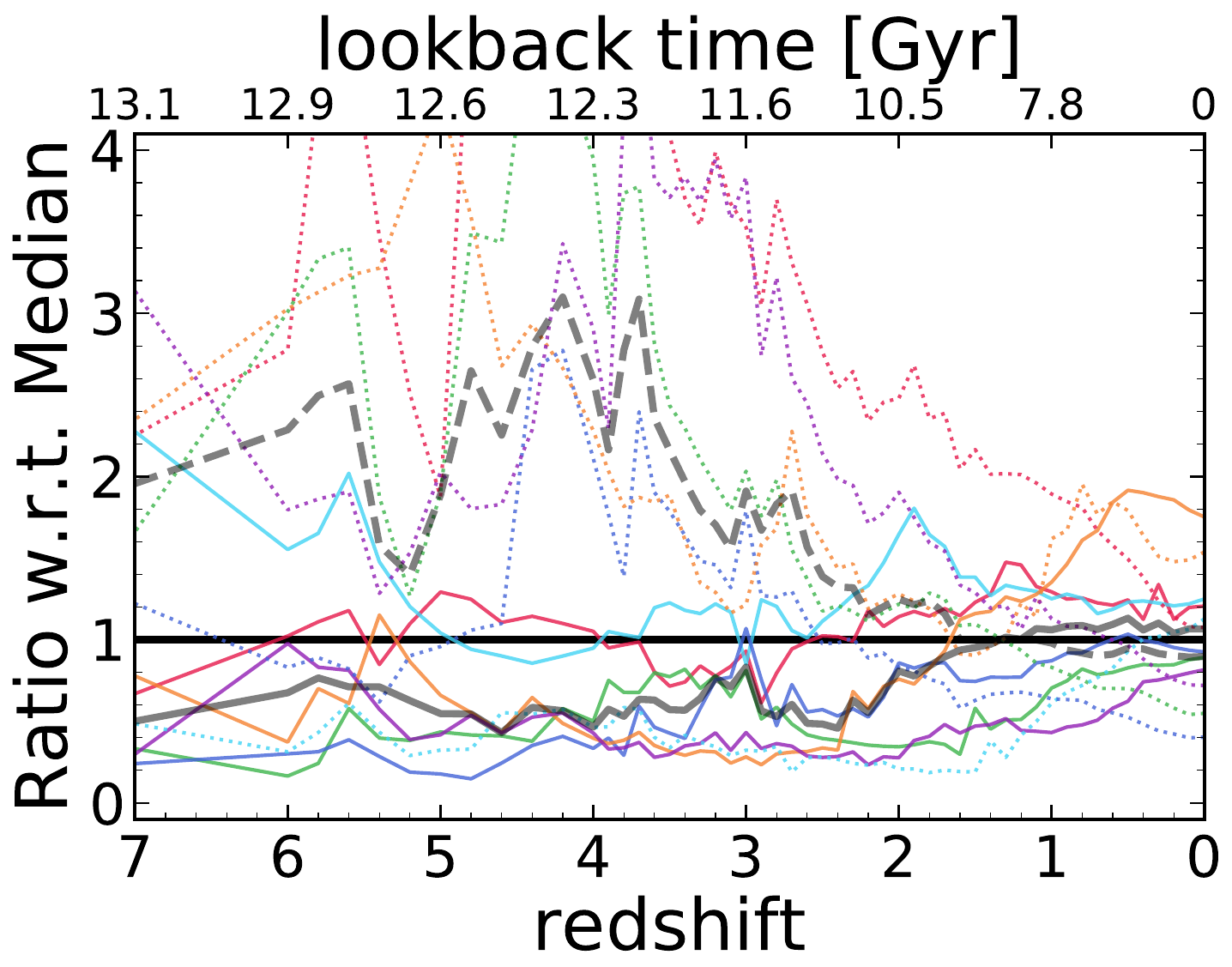} \\
\includegraphics[width=0.49\linewidth]{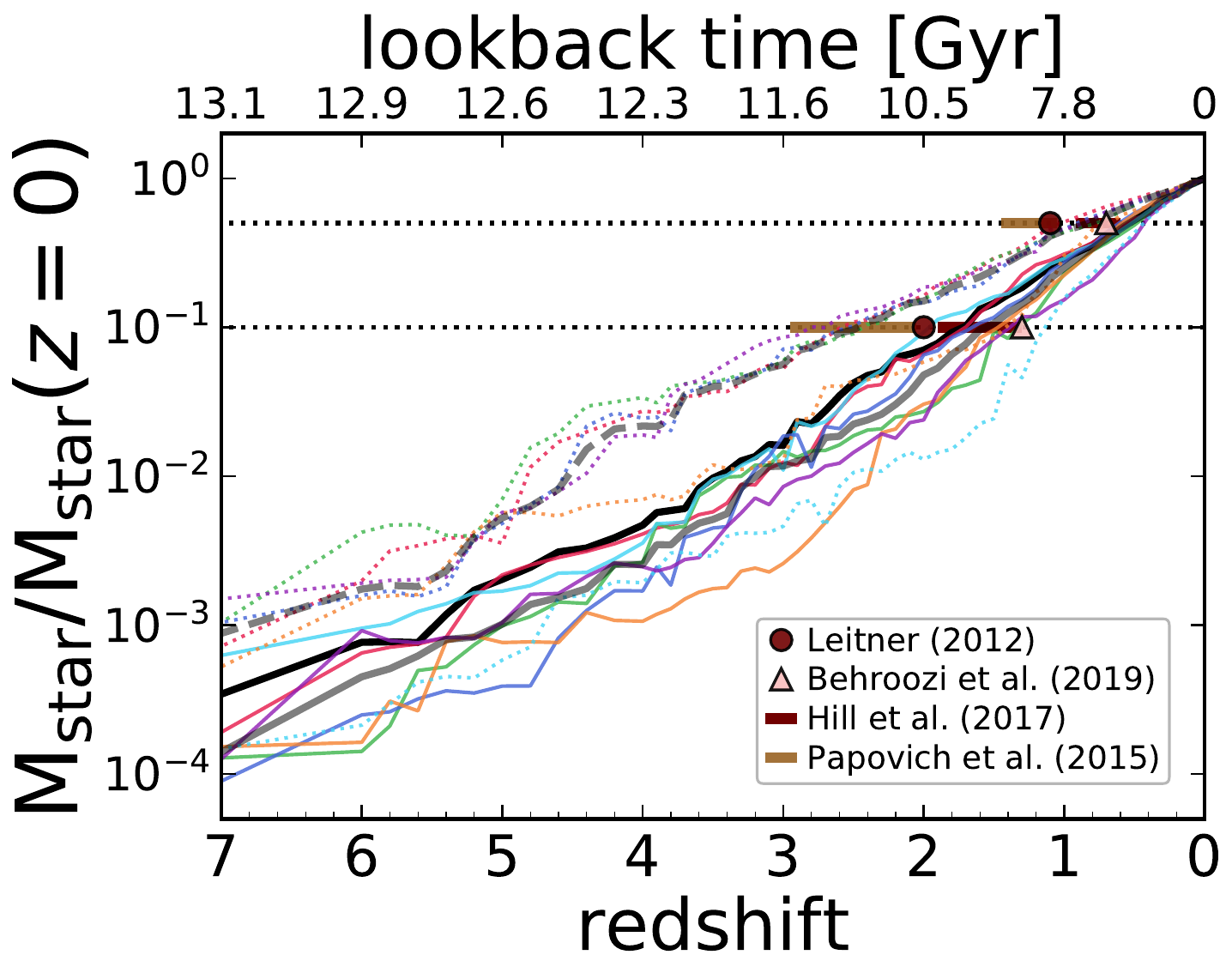} &
\includegraphics[width=0.49\linewidth]{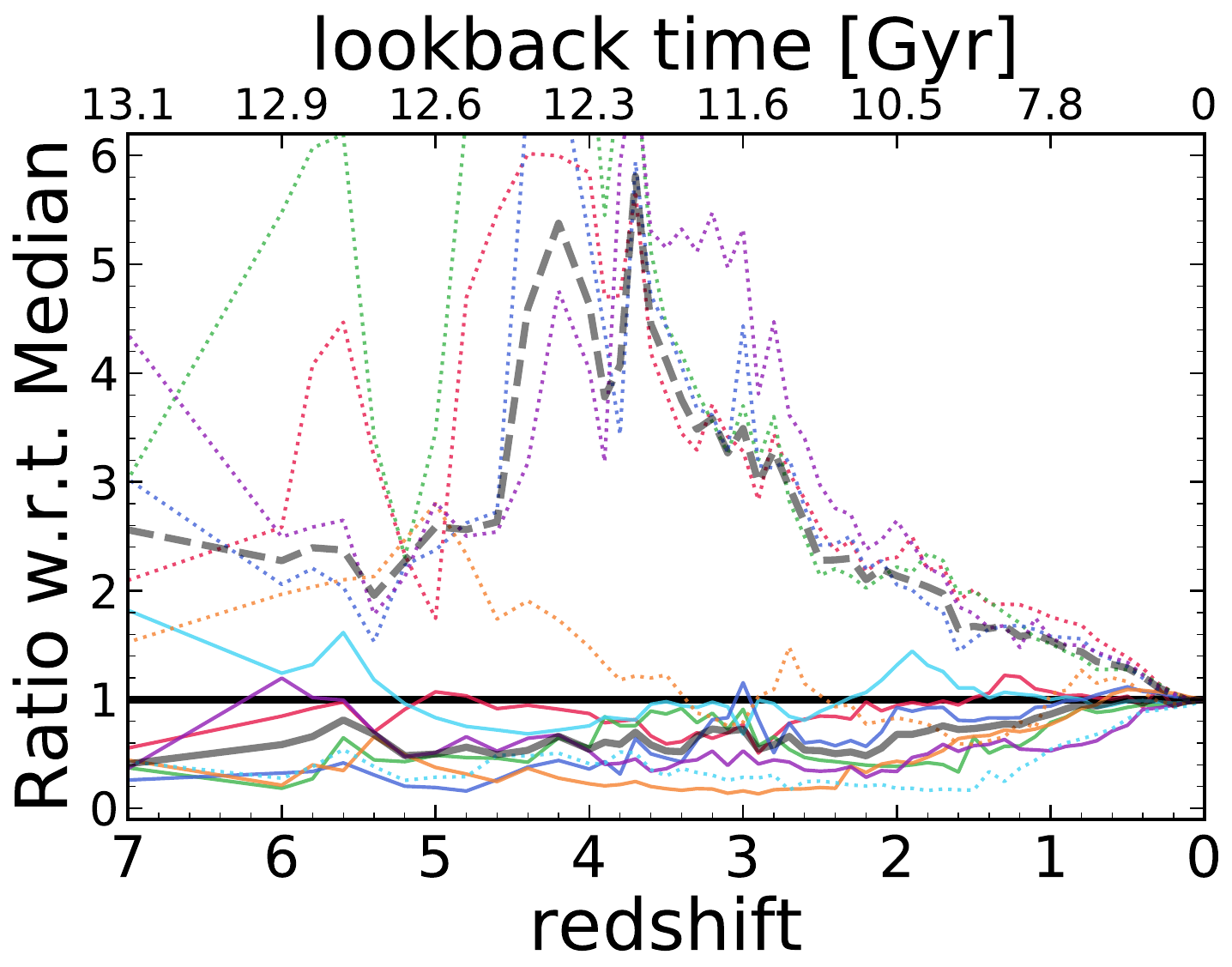}
\end{tabular}
\vspace{-3 mm}
\caption{
\textbf{Top Left:} The stellar mass of the most massive progenitor (MMP) of each MW/M31-mass galaxy as a function of redshift (bottom axis) or lookback time (top axis). 
Thin colored lines show each simulation, including the 6 isolated hosts (solid) and 6 LG-like hosts (dotted).
We also show the median $\Mstar$ across all 12 hosts (thick solid black), and the median for the isolated (thick solid grey), and the LG-like (thick dashed grey) hosts.
At $z = 0$ the host galaxies span a relatively narrow range of $\Mstar(z = 0) = 2.3 - 10 \times 10^{10} \Msun$ (by selection), but because of scatter in formation history, their MMPs spanned about 1-1.5 orders of magnitude at all $z \gtrsim 4$ ($> 12.2$ Gyr ago), with a range of $6 \times 10^{6} - 7 \times 10^{7} \Msun$ at $z = 7$, near the end of cosmic reionization.
\textbf{Top Right:} Same as left but normalized to the median $\Mstar(z)$ across all simulations.
Romeo reached the highest value of $\sim 6.5$ at $z = 4 - 5$ ($12.2 - 12.6$ Gyr ago), while m12i was the lowest at $\sim 0.15$.
\textit{The MMPs of LG-like paired hosts had significantly higher mass on average, by up to a factor of 6, at all $z \gtrsim 2$.}
\textbf{Bottom Left:} Same as top left but showing \textit{fractional} mass growth, with each MMP normalized by its $\Mstar(z = 0)$. The dotted horizontal lines indicate 10 per cent and 50 per cent of final mass.
Considering the overall median, the MMP reached 10 per cent of its final mass by $z \sim 1.7$ (9.9 Gyr ago), while for isolated hosts it was $z \sim 1.5$ (9.4 Gyr ago) and for LG-like pairs it was $z \sim 2.4$ (11.0 Gyr ago).
The MMP reached 50 per cent of its final mass by $z \sim 0.5$ (5.1 Gyr ago) across all hosts (and isolated host), and while this occurred at $z \sim 0.8$ (6.9 Gyr ago) for LG-like hosts.
We also plot observational inferences for galaxies that span our mass range at $z = 0$ (see text).
\textbf{Bottom Right:} Same as bottom left, but normalized to the total median: $\left[ \Mstar(z)/\Mstar(z=0) \right] / \left[ \Mstar(z)/\Mstar(z=0) \right]_{\rm med}$.
Here, the enhancement for LG-like hosts is even most dramatic, \textit{persisting at all redshifts and being almost $10 \times$ that of isolated hosts at $z \sim 4.2$} (12.3 Gyr ago).
This highlights the importance of environment in MW/M31-mass galaxy formation, especially at early times.
}
\label{fig:mmp}
\end{figure*}

Fig.~\ref{fig:mmp} (top left) shows the stellar mass of the MMP versus redshift (bottom axis) and lookback time (top axis).
We show each simulation individually (colored lines), including 6 isolated hosts (solid) and 6 LG-like hosts (dotted).
We also show the median across all 12 hosts (thick solid black), and the median for the isolated (thick solid grey) and the LG-like (thick dashed grey) hosts.
For clarity, we define $\Mstar(z)$ as the stellar mass inside of the MMP galaxy that has formed up to redshift $z$.
At $z = 0$ the host galaxies span a relatively narrow range of $\Mstar(z = 0) = 2.3 - 10 \times 10^{10} \Msun$ (by selection).
The shapes of the MMP mass growth histories show broad similarities, with $\log \Mstar$ growing almost linearly with redshift.
However, because of scatter in formation history, the $\Mstar$ of MMPs spanned 1-1.5 orders of magnitude at $z \gtrsim 2$ ($> 10.5 \Gyr$ ago).
For example, the MMPs spanned a range of $5.6 \times 10^{6} - 7.3 \times 10^{7} \Msun$ at $z = 7$, near the end of cosmic reionization.

Fig.~\ref{fig:mmp} (top right) shows the same, except we normalize each MMP to the overall median $\Mstar(z)$, which more clearly highlights both scatter in mass growth and systematic differences between LG-like and isolated hosts.
In particular, the typical $\Mstar$ of the MMP of an isolated host is similar (slightly higher) than that of a LG-like host at all $z < 1.5$, but prior to this, the MMP of an LG-like host was significantly more massive than that of an isolated host.
The biggest difference occurred at $z \sim 4$ ($\sim 12.2$ Gyr ago), when LG-like hosts were $\sim 6 \times$ more massive than isolated hosts.

Fig.~\ref{fig:mmp} (bottom left) shows the \textit{fractional} $\Mstar$ growth of each MMP, that is, $\Mstar(z)/\Mstar(z = 0)$.
This metric is more fair to compare the formation histories, because it normalizes out the scatter in $\Mstar(z = 0)$.
Across all simulations, the MMP reached 10 per cent and 50 per cent of its final mass at typically $z = 1.7$ (9.9 Gyr ago) and $z = 0.5$ (5.1 Gyr ago), respectively.
Again, the MMP of an isolated host typically formed later, reaching 10 per cent and 50 per cent of its final mass at $z = 1.5$ (9.4 Gyr ago) and $z = 0.5$ (5.1 Gyr ago), while the MMP of a LG-like host typically formed earlier, reaching 10 per cent and 50 per cent of its final mass at $z = 2.4$ (11.0 Gyr ago) and $z = 0.8$ (6.9 Gyr ago).

Finally, Fig.~\ref{fig:mmp} (bottom right) shows the fractional $\Mstar$ growth in the bottom left panel but normalized to the total median (black curve) across all simulations.
This metric shows that, by normalizing the curves by their $\Mstar$ at present, the median LG-like host now had higher fractional mass than the median isolated host \textit{at all redshifts}.
The biggest difference occurred at $z \sim 4.2$ ($\sim 12.3$ Gyr ago), where the medians differed by almost an order of magnitude.

While this offset between LG-like and isolated hosts has some host-to-host scatter, we emphasize that at \textit{all} redshifts, the MMP for 4 or 5 of the 6 LG-like hosts was fractionally more massive than the MMP for \textit{all} of the isolated hosts.
This result highlights significant systematic differences in the formation histories of the MMPs in LG-like versus isolated environments.
This is important when using cosmological zoom-in simulations such as these to compare with observations of the MW and M31 in the LG or with isolated MW-mass hosts as in, for example, the SAGA survey.
We also emphasize that the MMPs of LG-like hosts were systematically more massive even back to $z = 7$, indicating these environmental effects manifest themselves early, even when the MMPs were only $10^{-4}$ to $10^{-3}$ of their final stellar mass.
In Section~\ref{sec:mmphalo}, we discuss how this likely follows from the mass growth of the DM host halo.

These trends are consistent with a similar analysis of the archaeological star-formation histories (SFHs) of these hosts in \citet{GarrisonKimmel19a}, which calculated SFHs using all stars in the host galaxy at $z = 0$.
By selecting stars in the host galaxy at $z = 0$, and using their formation times alone to calculate the SFH of the host galaxy, this ultimately includes stars that originally could have formed in a galaxy other than the MMP.
This is different from how we compute the mass growth curves in Fig.~\ref{fig:mmp}, where we compute the $\Mstar$ of the MMP at each snapshot.
Therefore, this result is not sensitive to the way that one computes the stellar mass growth of the host galaxy.

For comparison, the colored points/bands in Fig.~\ref{fig:mmp} (bottom left) show observational inferences of $\Mstar(z)$ in MW/M31-mass galaxies.
Using a fit for the SFR main sequence to observational data of star-forming galaxies, \citet[][red circles]{Leitner12} constructed mass growth histories, finding that galaxies of $\Mstar(z = 0) \sim 10^{11}$ reached 10 per cent of their final mass at $z \approx 2$ (10.4 Gyr ago), and 50 per cent of their final mass at $z \sim 1.1$ (8.2 Gyr ago).
These values are broadly consistent with our LG-like hosts, though they are earlier than our typical isolated host, but our hosts have somewhat lower  $\Mstar(z = 0)$ than in \citet{Leitner12}, so this may not be surprising.
In another study, \citet{Papovich15} used abundance matching to find progenitors of MW/M31-mass galaxies, and calculated their stellar mass evolution, taking into account mass loss from stellar evolution.
Galaxies with $\Mstar(z = 0) = 5\times10^{10} \Msun$ and $\Mstar(z = 0) = 10^{11} \Msun$ (akin to the MW and M31) tend to reach 10 per cent of their final mass around $z \sim 2 - 3$ and 50 per cent around $z \sim 1 - 1.5$, respectively.
We represent these redshift ranges via the light brown bands in the bottom left panel and note that these results are consistent with the curves for our LG-like hosts.
More recently, \cite{Hill17} also used abundance matching of stellar mass functions at various redshifts to track MW/M31-mass growth histories \citep[e.g.][]{vanDokkum13}.
\cite{Hill17} found 10 per cent mass occurred at $z \sim 1.3 - 1.9$ (8.9 - 10.2 Gyr ago) for $10^{10.5 - 11} \Msun$ galaxies, and 50 per cent mass occurred at $z \sim 0.6 - 0.9$ (5.8 - 7.4 Gyr ago).
We show this range (which brackets our simulation sample) via two dark red bands in Fig.~\ref{fig:mmp}, which also broadly agree with our simulation suite, though we note that the typical 50 per cent mass in our simulations occurs slightly later than all of these observational estimates.
Finally, \citet{Behroozi19} (pink triangles) used the Bolshoi-Planck DM-only (DMO) simulation, combined with observed properties of galaxies such as stellar mass functions and SFRs to determine the SFHs of galaxies with $\Mthm = 10^{12} \Msun$ (similar to our sample).
They found that these galaxies reached 10 per cent and 50 per cent of their final $\Mstar$ around $z \approx 1.3$ and $z \approx 0.7$ ($\approx 8.9$ and $\approx 6.4$ Gyr ago, respectively), which are consistent with the typical values in our simulations.

\subsection{What were the building blocks of the galaxy?}
\label{sec:mfs}

\begin{figure*}
\centering
\begin{tabular}{c @{\hspace{-0.1ex}} c}
\includegraphics[width=0.45\linewidth]{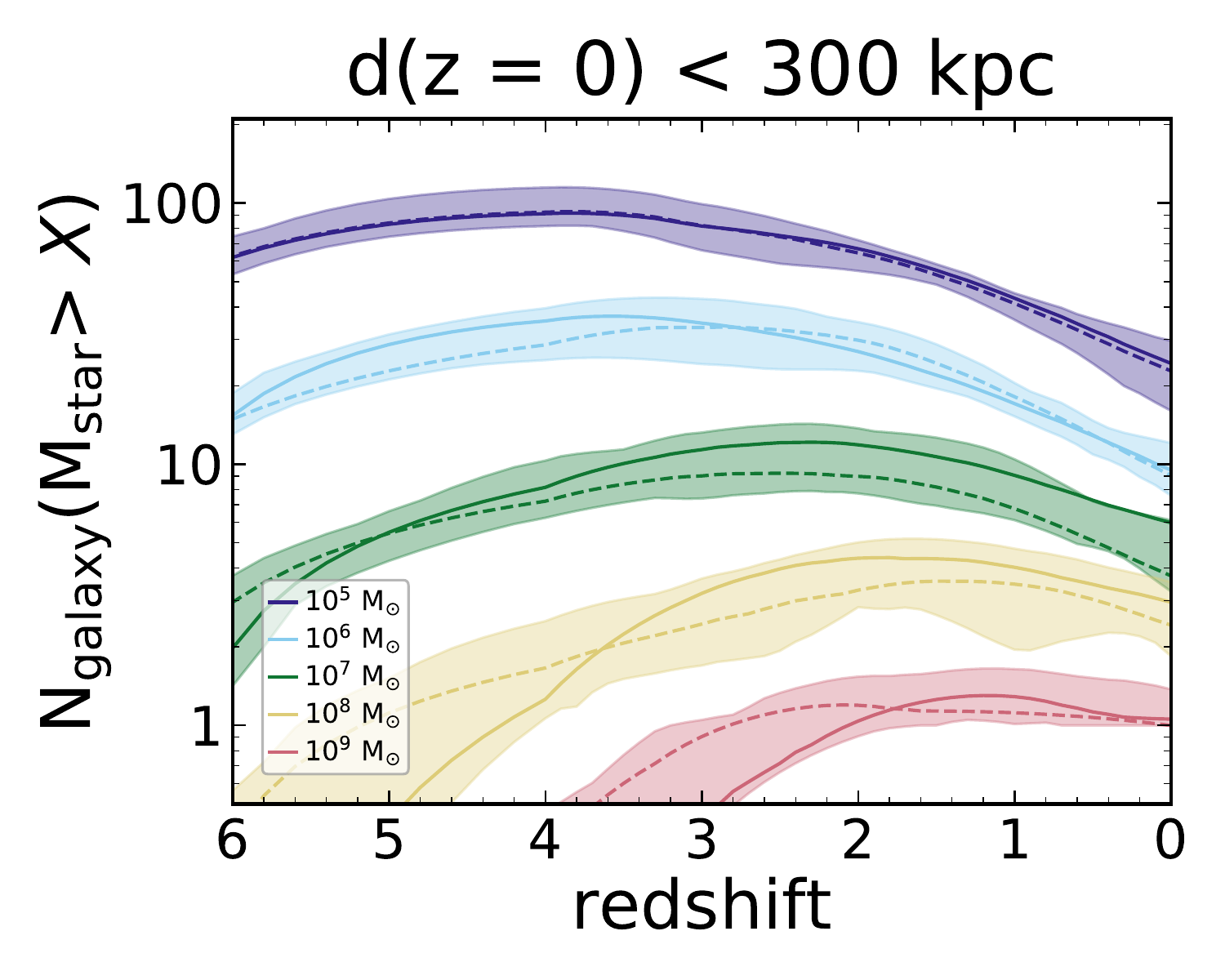}&
\includegraphics[width=0.45\linewidth]{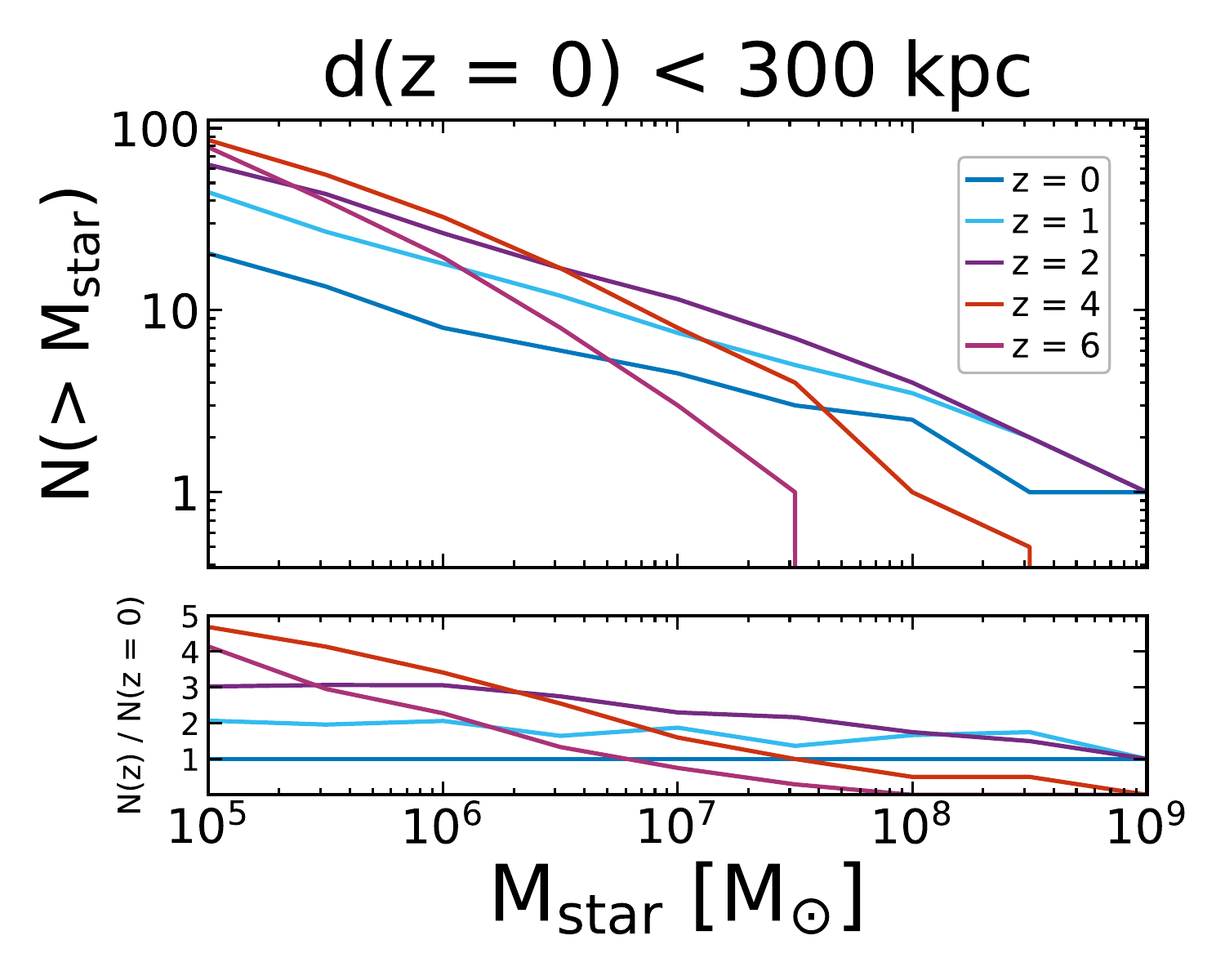} \\
\includegraphics[width=0.45\linewidth]{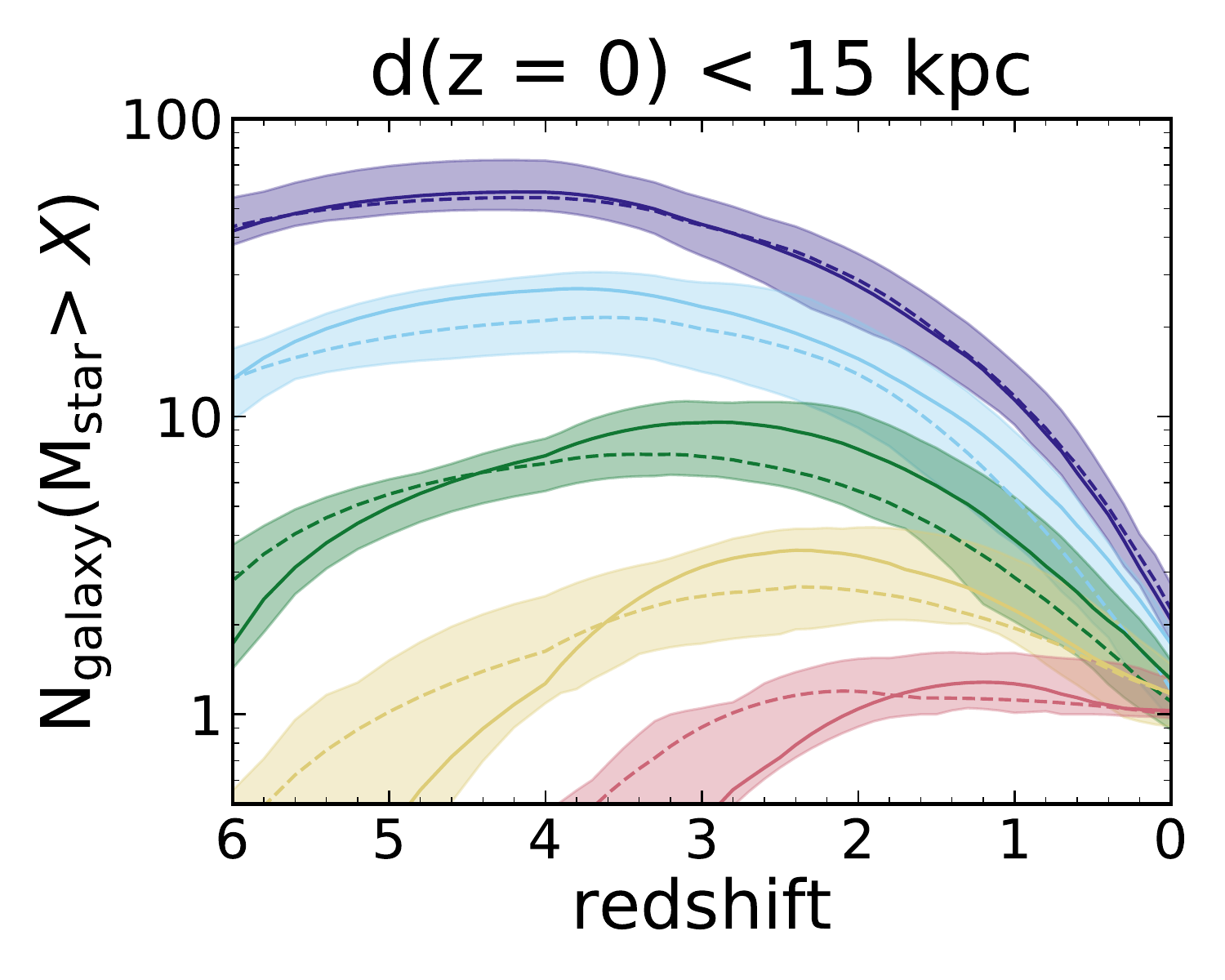} &
\includegraphics[width=0.45\linewidth]{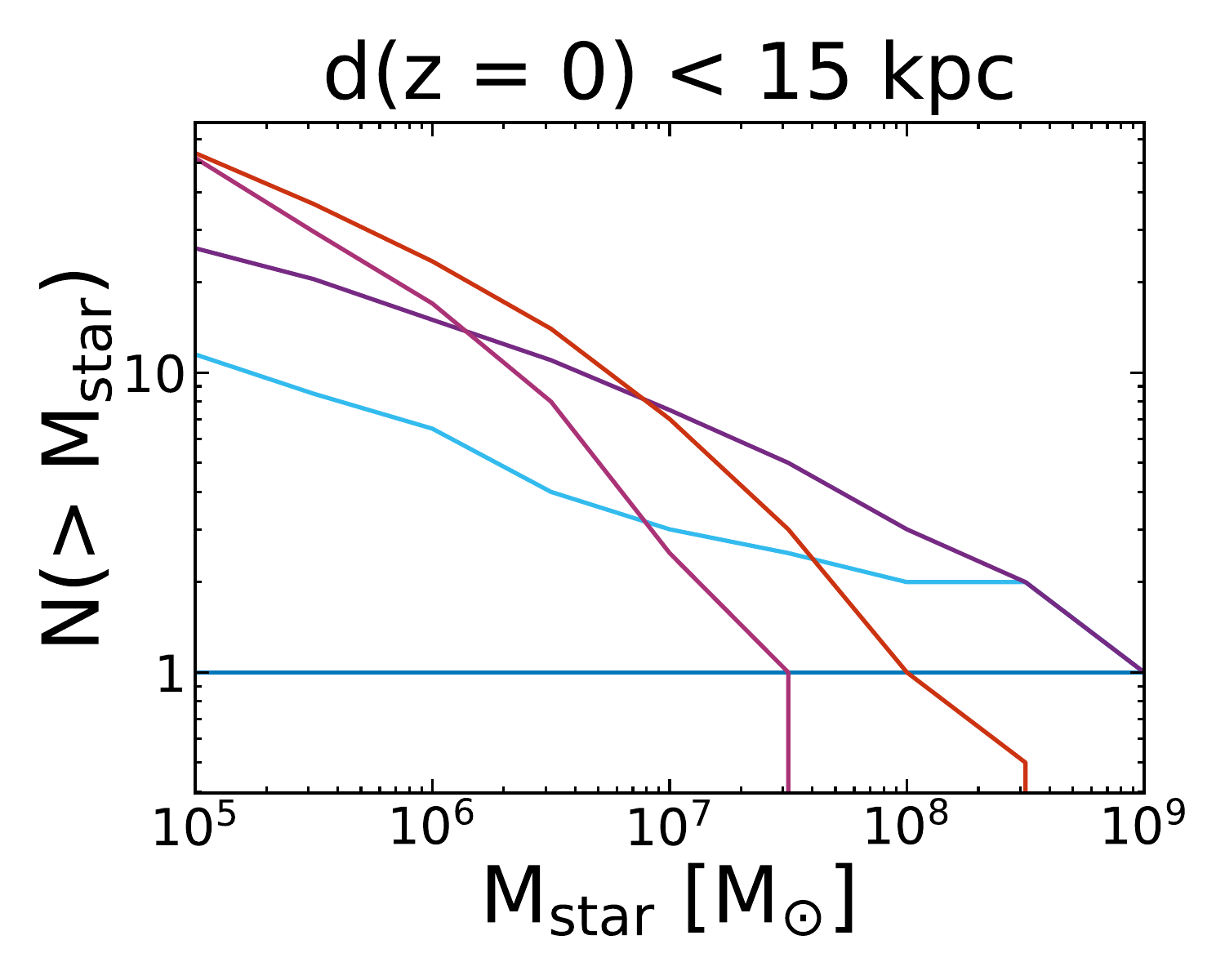} \\
\includegraphics[width=0.45\linewidth]{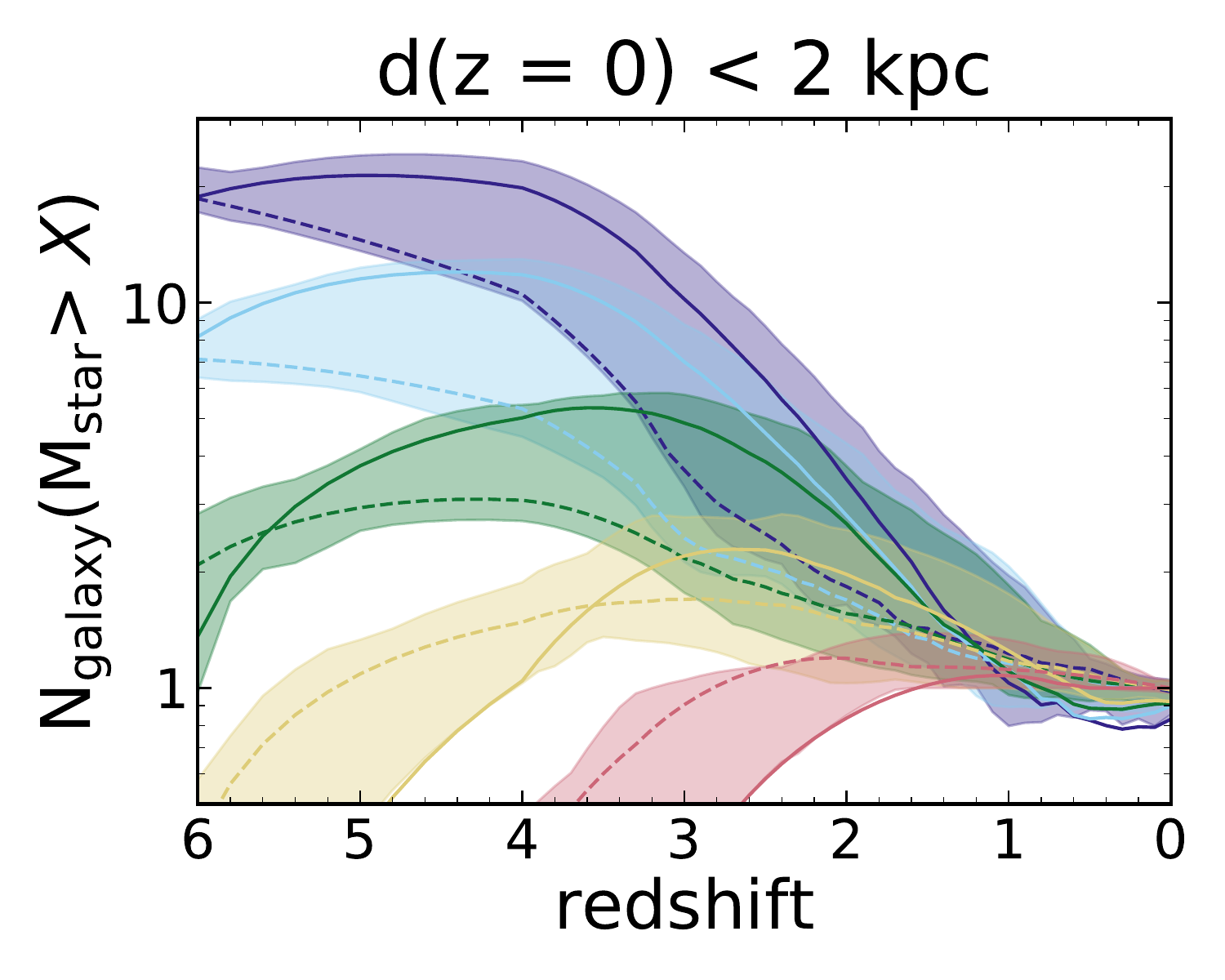} &
\includegraphics[width=0.45\linewidth]{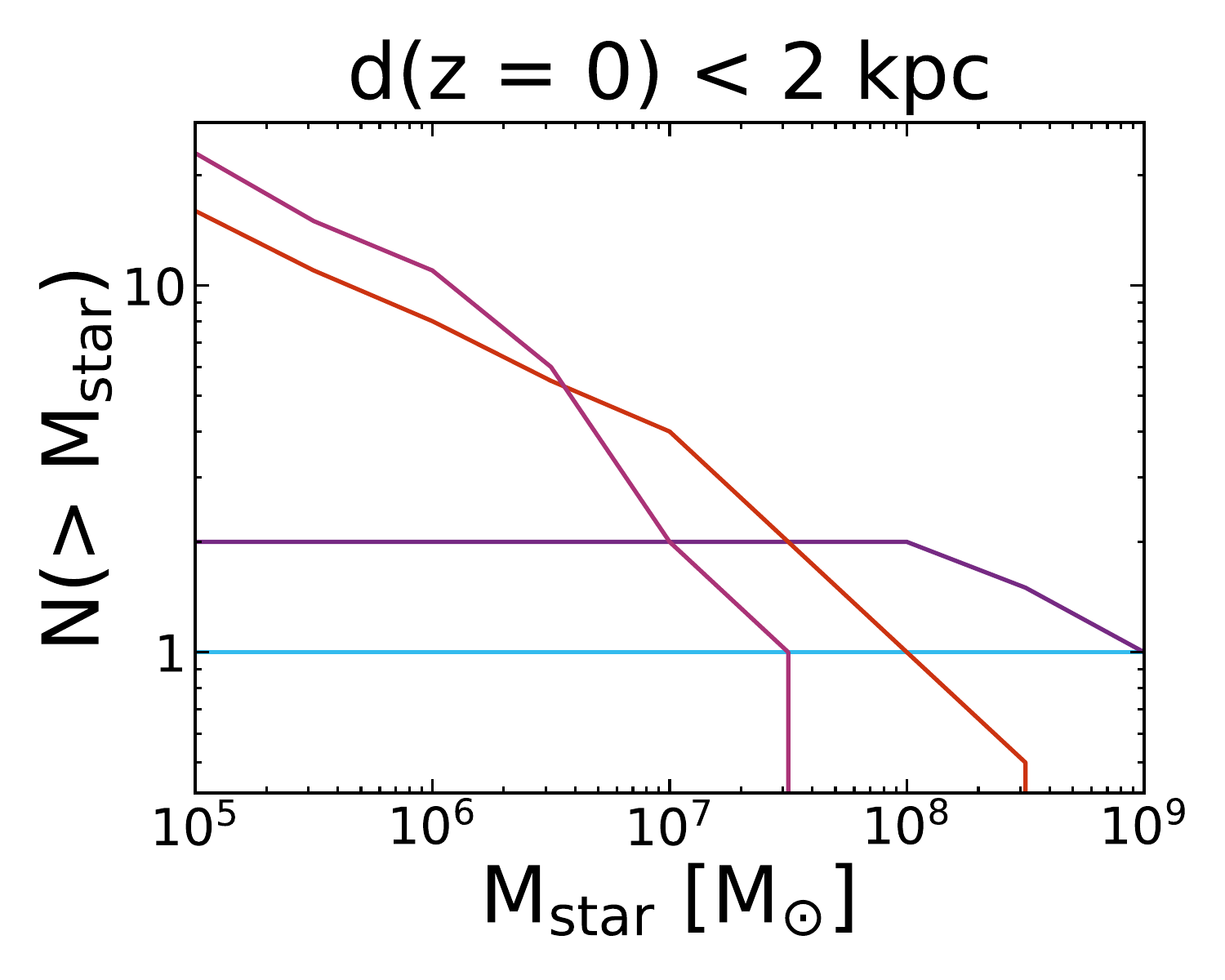}
\end{tabular}
\vspace{-2 mm}
\caption{
\textbf{Left:} The number of progenitor galaxies above a given stellar mass versus redshift.
Progenitor galaxies are those whose stars end up within the given host-centric distances at $z = 0$: the entire host halo (top), the host galaxy + inner stellar halo (middle), and the inner bulge region (bottom).
The lines show the median for our 6 isolated (solid) and 6 LG-like (dashed) hosts, while the shaded regions show the 68 per cent scatter across all 12 hosts.
Both LG-like and isolated hosts show the same general trends, though the progenitor population of isolated hosts peaks slightly later.
\textbf{Right:} Cumulative number of progenitor galaxies versus stellar mass at different redshifts, for the median across all 12 hosts.
For $d_{300}$ (top), the lower panel shows the number at each redshift normalized to the number at $z = 0$, that is, the population of progenitors relative to surviving satellites at $z = 0$.
Note that the progenitor mass function becomes increasingly steeper at higher redshifts (for more discussion see Section~\ref{sec:disc2}).
Higher-mass progenitors contributed preferentially at later times.
For $\Mstar > 10^5 \Msun$, $\sim$ 85-100 progenitor galaxies contributed to the formation of a MW/M31-mass system out to its halo virial radius, $\sim$ 55 contributed to the galaxy within 15 kpc, and $\sim$ 20 contributed to the inner bulge region within 2 kpc, with the number of progenitors peaking at $z \sim 4$.
Thus, the current satellite population at $\Mstar > 10^5 \Msun$ represents only $\sim 1 / 5$ of the population that formed each system.
}
\label{fig:mfs}
\end{figure*}

We next investigate the distribution of `building blocks' of our MW/M31-mass host galaxies by analyzing the cumulative stellar mass function (number of galaxies above a $\Mstar$ threshold) of all progenitor galaxies (\textit{including} the MMP) across time.
We select progenitors that contribute to each of our host-centric distance cuts at $z = 0$.
In Fig.~\ref{fig:mfs}, the left panels show the median number of progenitor galaxies for the isolated (solid lines) and LG-like (dashed lines) hosts versus redshift using $\Mstar$ thresholds.
The shaded regions show 68 per cent scatters across our total sample.
The panels on the right show the median number of progenitors across the total sample versus $\Mstar$ at given redshifts (we do not show isolated and LG-like hosts separately here).

Fig.~\ref{fig:mfs} highlights several interesting trends.
For the lowest-mass progenitors that we resolve ($\Mstar > 10^5 \Msun$), the number of progenitor galaxies peaked at $z \sim 4 - 5$ ($12.2 - 12.6$ Gyr ago) for all host distance selections.
For $d_{300}$, the number of progenitors peaked at $\sim 85-100$, while for $d_{15}$ it peaked at $\sim 55$, and for $d_2$ it peaked at $\sim 20$.
Given hierarchical structure formation, there were fewer higher-mass progenitors, and their numbers peaked at progressively later times.
For example, for $d_{300}$, the number of progenitors with $\Mstar > 10^7 \Msun$ peaked at $\sim 10$ at $z \sim 2.5$.
We see similar trends for 15 and 2 kpc distance cuts, though the progenitor peaks shift to larger redshifts, indicating that the increasingly central regions of the MMP formed earlier.
For our highest stellar mass cut, $\Mstar > 10^9 \Msun$, the hosts typically have only one progenitor (the MMP itself).
However, both the MW (with the LMC) and M31 (with M32 and M33) have 1-2 satellites above this mass today.
While none of our hosts possess such massive satellites at $z = 0$, half of our hosts (6 of 12) have had at least one satellite with $\Mstar > 10^9 \Msun$ since $z = 0.7$ (Chapman et al., in prep.), but these massive satellites merge into the host galaxy quickly because of efficient dynamical friction, resulting in the instantaneous median number being 0 across all simulations.

The top right panel also shows the number of progenitors at each redshift normalized to the surviving satellites within 300 kpc at $z = 0$.
Compared to the present satellite galaxy population, there were nearly $\sim 5$ times as many dwarf galaxies that formed each entire MW/M31-mass system, with their numbers peaking at $z \sim 4$ (12.2 Gyr ago).
Most of these low-mass progenitors have become disrupted into the main galaxy today.
This has implications for galactic archaeology and the ``ear-far' connection, specifically for any attempt to use present-day nearby dwarf galaxies to make inferences about the high-redshift universe (see Section~\ref{sec:disc2}).
Note that the bottom right two panels show just one galaxy (the host galaxy) at $z \sim 0$, because typically these hosts have no resolved satellites inside of 15 kpc \citep{Samuel19}.
For $d_{15}$, typically 10 progenitor galaxies have contributed to its formation since $z = 1$ (7.8 Gyr ago), while typically only 1-2 progenitors other than the MMP have contributed to the bulge region at $z \lesssim 2$.

We note that for all host distance cuts, the mass function of progenitors was increasingly steeper at higher redshifts which is consistent in both observational and simulation results of the general galaxy population \citep[e.g.][see Section~\ref{sec:disc2} for more discussion]{Graus16, Song16, Ma18}.

Finally, the satellite populations of isolated and LG-like hosts do not differ substantially across time, as seen in the left panels of Fig.~\ref{fig:mfs}.
For mass bins with $\Mstar \geq 10^7 \Msun$, LG-like hosts initially had more progenitors, reflective of their earlier formation histories, but are soon outnumbered compared to the isolated hosts.
These differences, however, are small and the right panels show the same behaviour, so we present only the total median in those cases.

\begin{figure*}
\centering
\begin{tabular}{c @{\hspace{-0.1ex}} c}
    \includegraphics[width=0.45\linewidth]{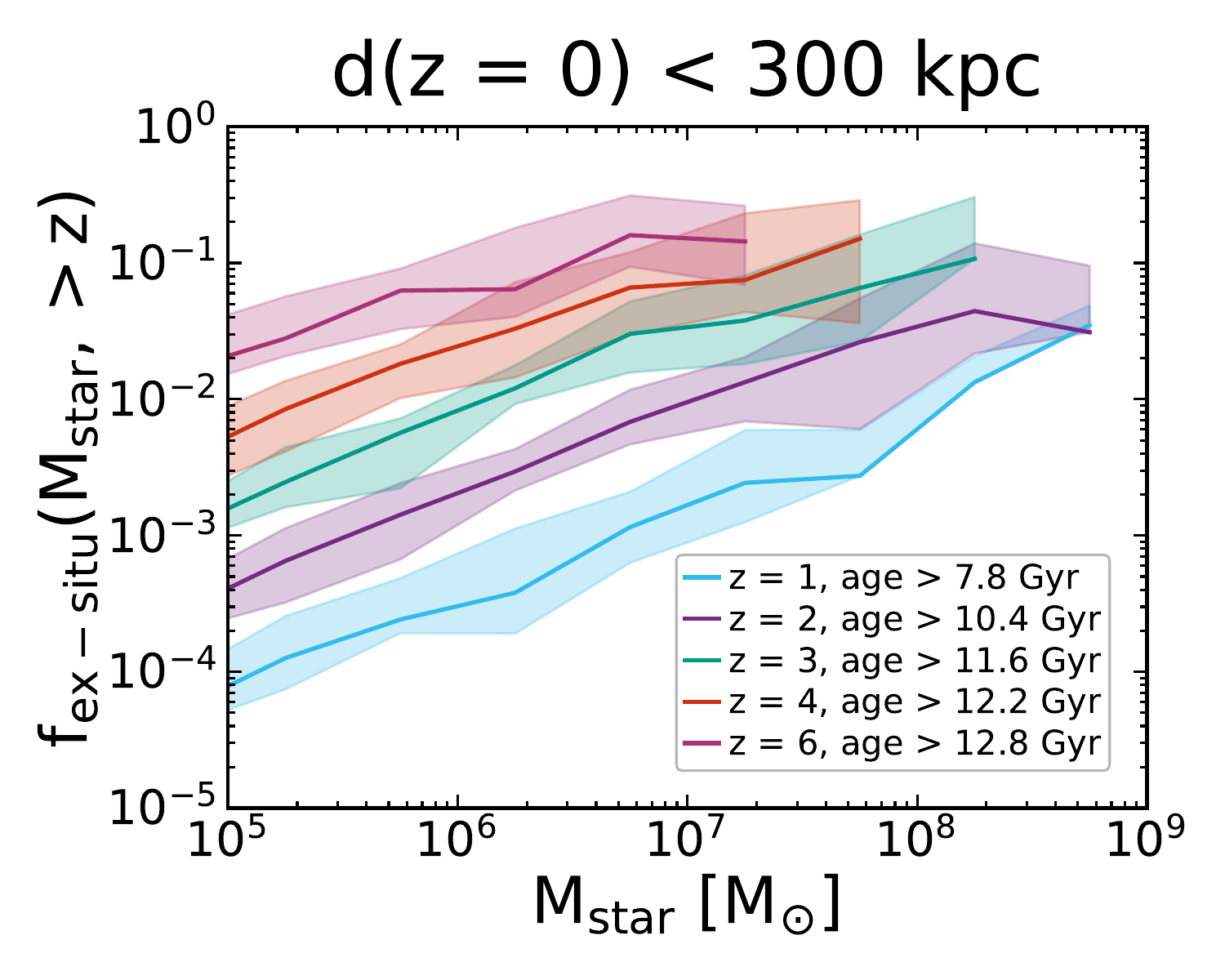}&
    \includegraphics[width=0.45\linewidth]{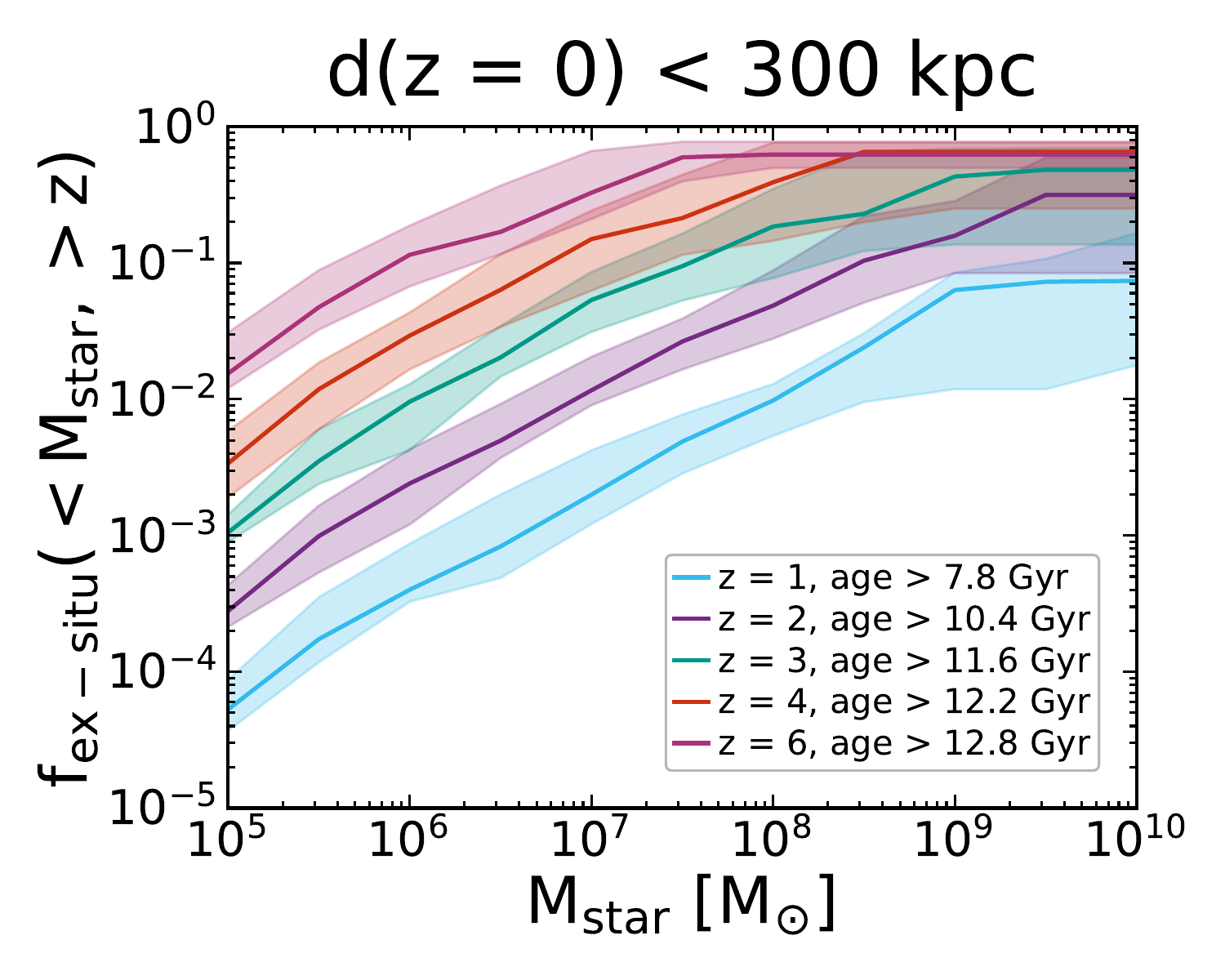}\\
    \includegraphics[width=0.45\linewidth]{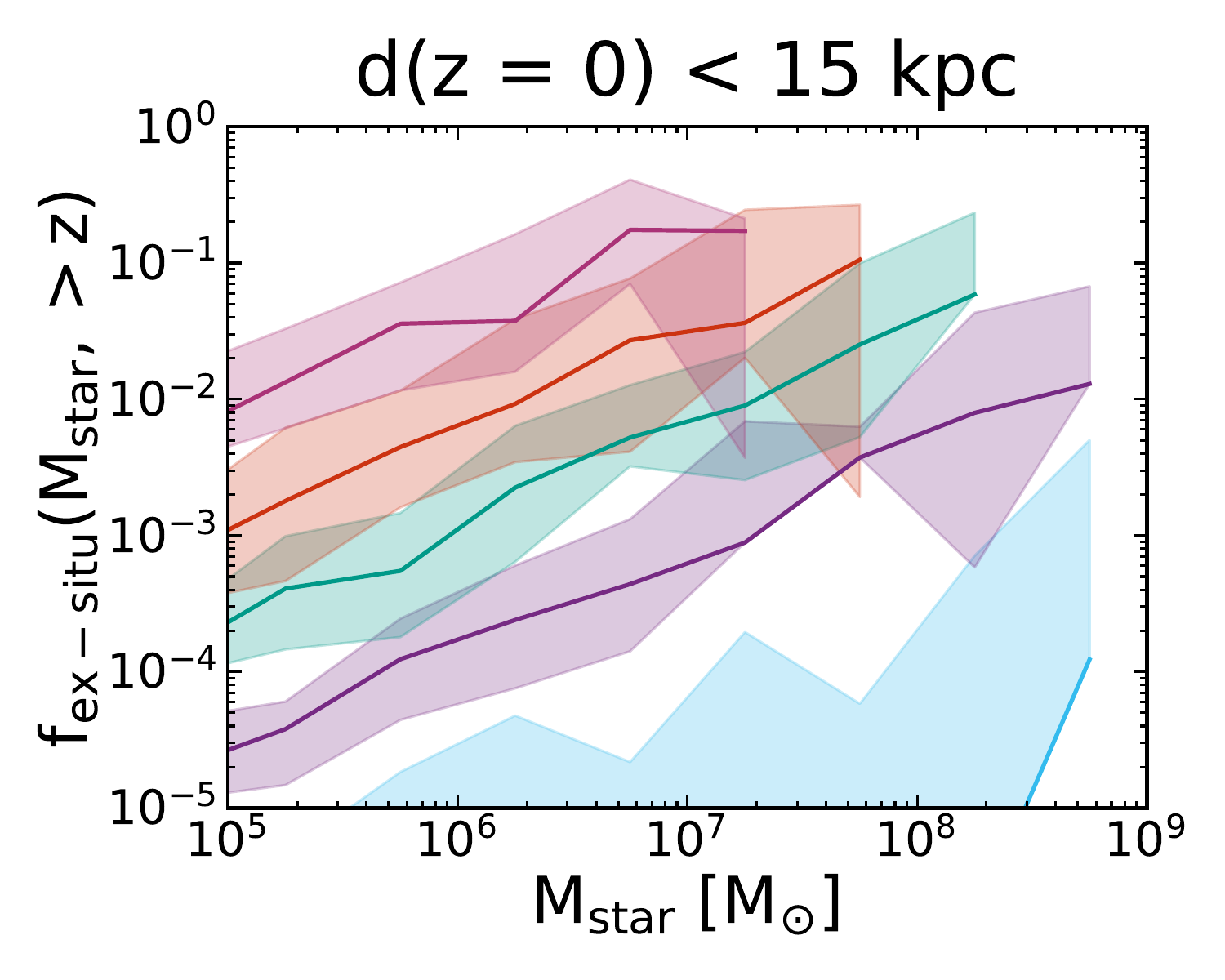}&
    \includegraphics[width=0.45\linewidth]{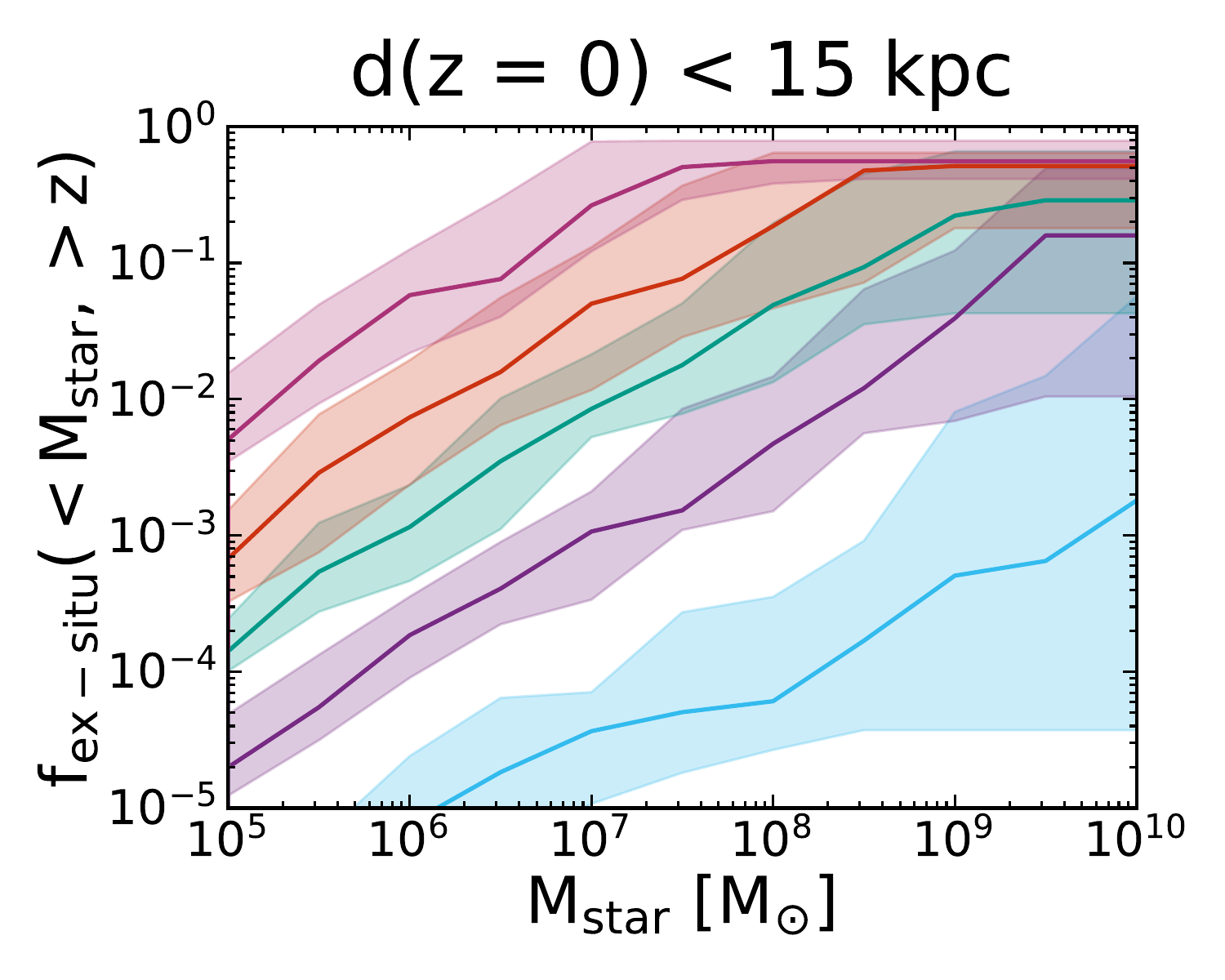}\\
    \includegraphics[width=0.45\linewidth]{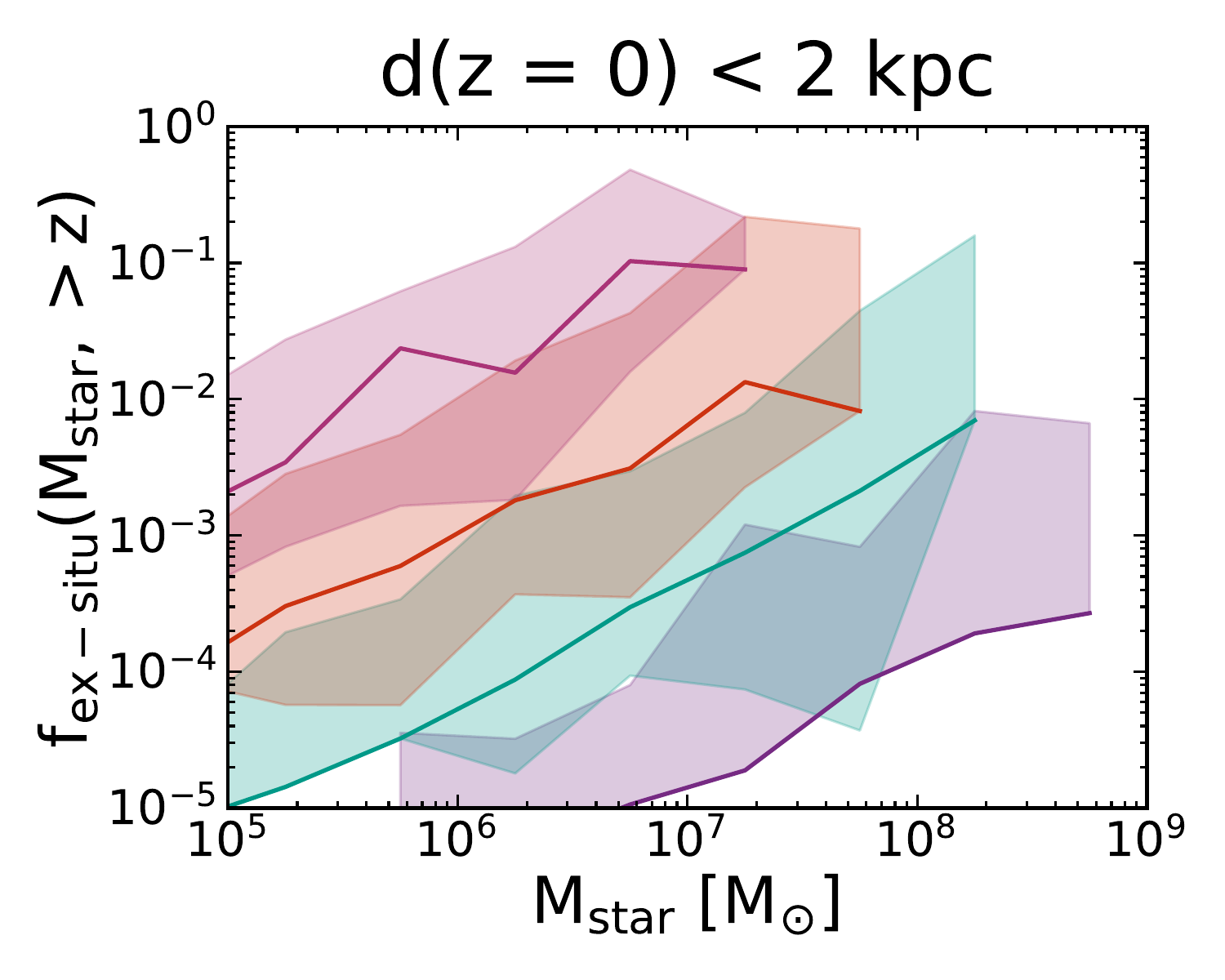}&
    \includegraphics[width=0.45\linewidth]{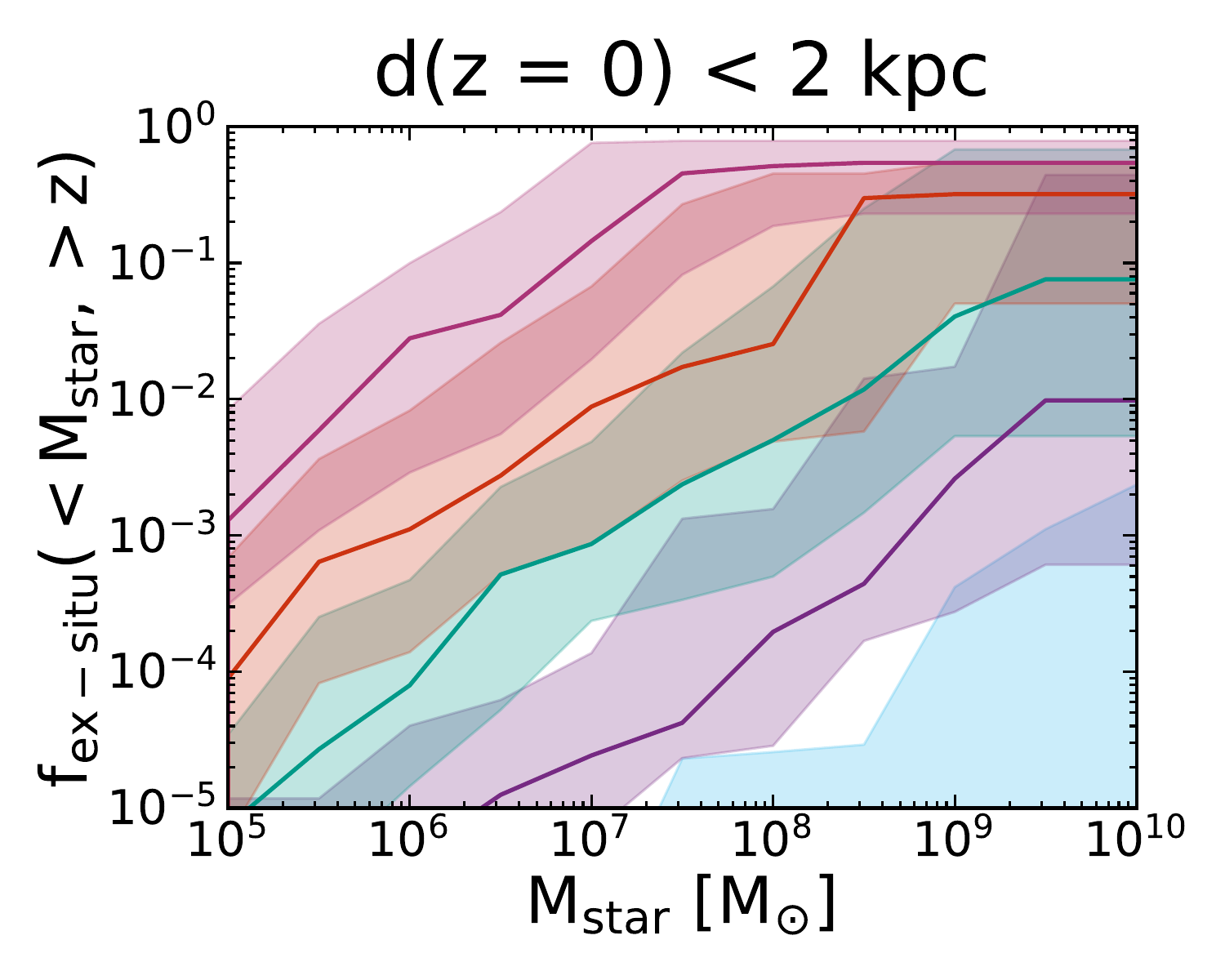}
\end{tabular}
\vspace{-2 mm}
\caption{
For all stars that end up within the given host-centric distances at $z = 0$ (top to bottom) and that formed before the given redshifts (different colored lines), panels show the fraction that formed ex-situ (in progenitors other than the MMP) as a function of progenitor stellar mass.
The left panels show the (differential) fraction within the $\Mstar$ bin, while the right panels show the cumulative fraction less than $\Mstar$.
Lines show total median while shaded regions show 68 per cent scatter across all 12 hosts (see text for discussion of the differences between isolated and LG-like hosts).
Because of our sample size, the lower scatter for some mass bins is 0: in these cases we set the lower scatter equal to the median for visual clarity.
For all host-centric distances, the ex-situ fraction increases monotonically with increasing redshift and with increasing progenitor mass, so more massive progenitors \textit{always} dominate the ex-situ mass.
}
\label{fig:wmfs}
\end{figure*}

Fig.~\ref{fig:mfs} shows that the lowest-mass progenitors dominated by number.
However, this does not mean that the lowest-mass progenitors dominated the ex-situ mass of the host (that is, the stellar mass that formed in progenitor galaxies other than the MMP).
To quantify this, at each redshift we take the population of progenitors in Fig.~\ref{fig:mfs} (right panels), excluding the MMP, and we weight each progenitor by their contribution fractions.
This gives us the stellar mass in these galaxies, that formed before this redshift, that eventually gets deposited into the host.
We then divide this quantity by the total stellar mass in the host at $z = 0$ (for the different distance selections $d_{300}$, $d_{15}$, \& $d_2$), to obtain the ex-situ fraction, $f_{\rm ex-situ}(\Mstar, >z)$; we show this in Fig.~\ref{fig:wmfs}.
This represents the fraction of stellar mass within the distance selections of the host at $z = 0$, that is above a given age and formed in progenitors of a given $\Mstar$ (other than the MMP).
Note that this does not equal the total fraction of $\Mstar$ formed ex-situ \textit{at all redshifts} within each host at $z = 0$ (see Fig.~\ref{fig:insitu} for that).
Instead, we analyze the ex-situ fraction for stars formed before a given redshift, to highlight trends if one selects stars of minimum age in the MW or M31 today.

Fig.~\ref{fig:wmfs} shows $f_{\rm ex-situ}(\Mstar, > z)$ versus progenitor $\Mstar$.
The curves in the left panels show the fraction of stellar mass (that formed before a given redshift) that ends up within the three host-centric distances at $z = 0$, that formed inside progenitors other than the MMP, as a function of progenitor stellar mass.
The right panels show the cumulative ex-situ fraction that formed in progenitors below a given stellar mass.
The shaded regions in all panels show the 68 per cent scatter across the simulations.
In some cases on the left panels, given our finite sample of simulations, the lower scatter goes to zero, that is, in some simulations no progenitors at that mass contributed any stars.
For visual clarity, we set those values for the lower scatter equal to the median.

Because the slope of the (differential) progenitor mass function, ${\rm d} N / {\rm d}\log M$, is shallower than unity, weighting progenitors by their $\Mstar$ means that more massive progenitors (other than the MMP) dominated the ex-situ mass, while the contributed $\Mstar$ from the lowest-mass galaxy progenitors was comparatively negligible.
Also, as a result of hierarchical structure formation, at later times increasingly more massive progenitors dominated the ex-situ mass.
For stellar populations of \textit{all} ages at $z = 0$, the ex-situ mass is dominated by the most massive progenitors, but that mass scale decreases with increasing age.

All of our host-centric distance cuts at $z = 0$ show the same trends above.
The primary difference is that regions closer to the center of the host at $z = 0$ have lower ex-situ fractions (normalizations) at all progenitor masses.
In other words, the ex-situ contribution becomes increasingly negligible towards the central regions of the host galaxy.

While we do not show it separately in Fig.~\ref{fig:wmfs} (see instead Fig.~\ref{fig:insitu}), we find that isolated hosts had larger ex-situ fractions on average than the LG-like hosts.
Specifically, the ex-situ fractions for isolated hosts were primarily at the upper end of the 68 per cent scatter region, and LG-like hosts were primarily near the lower end.
Thus, ex-situ growth is more important for isolated hosts, but any differences converge by $z \sim 1$.

\subsection{When did in-situ star formation dominate the mass growth?}
\label{sec:insitu}

\begin{figure*}
\centering
\begin{tabular}{c @{\hspace{-1ex}} c}
\includegraphics[width=0.49\linewidth]{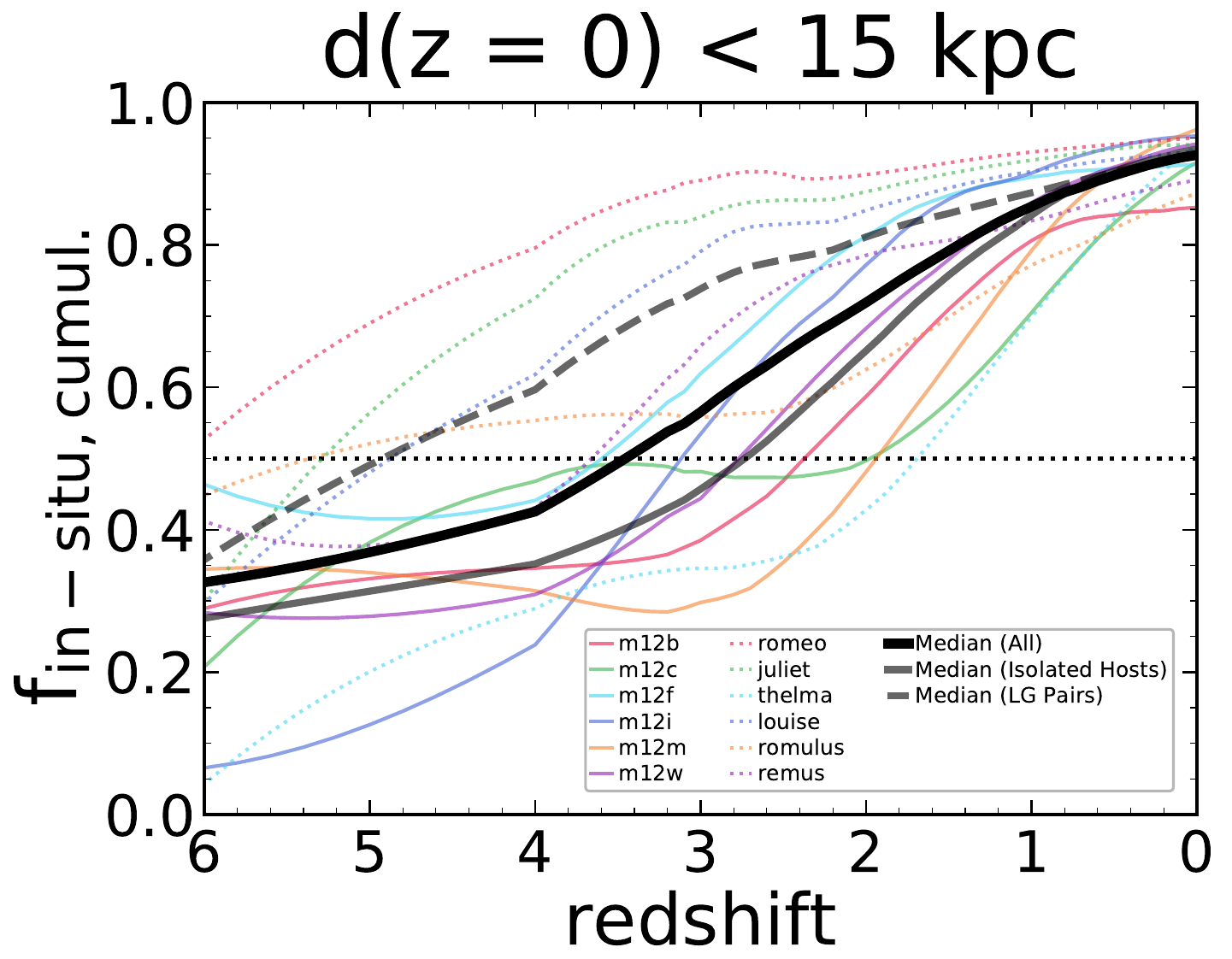}&
\includegraphics[width=0.49\linewidth]{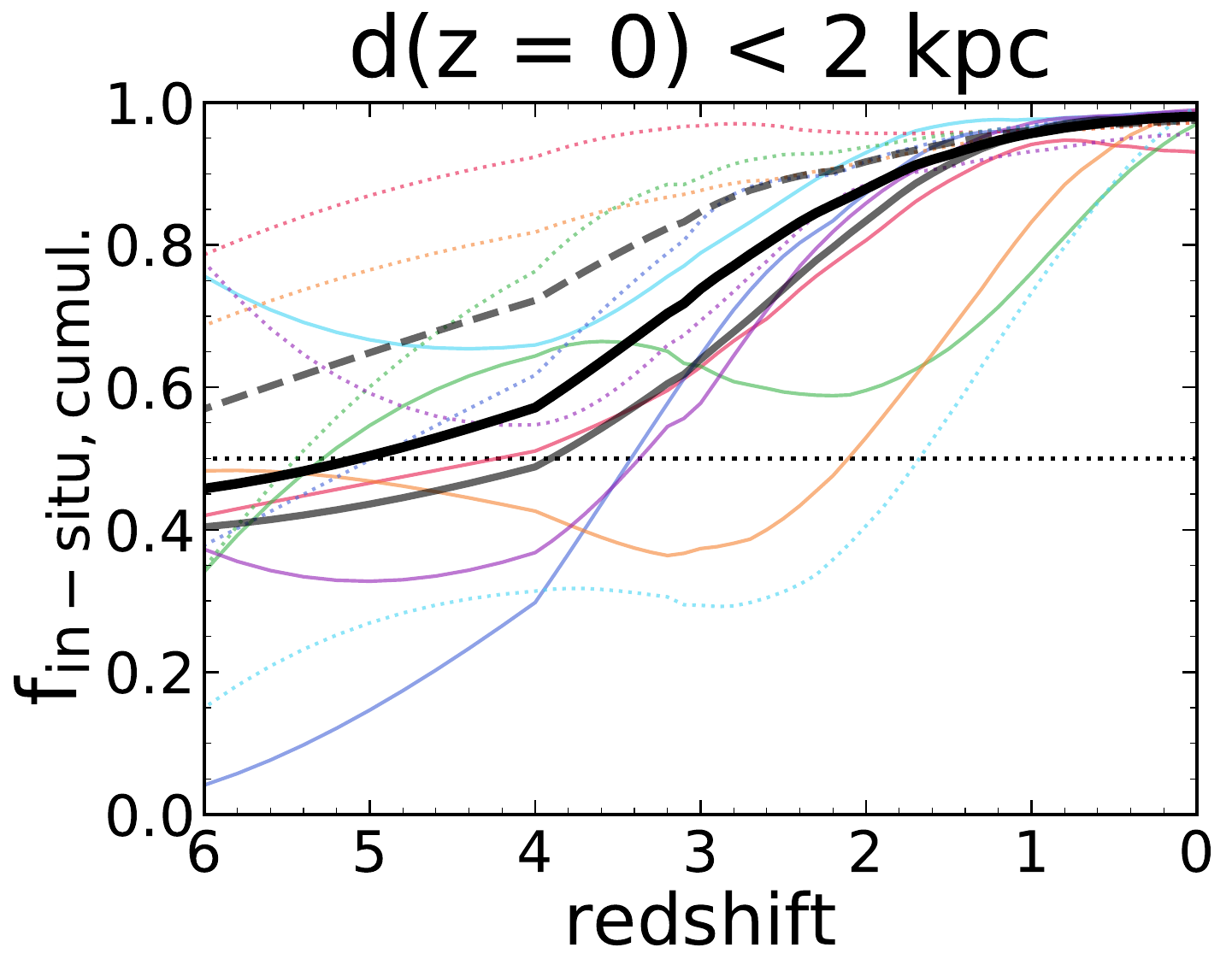}
\end{tabular}
\vspace{-2 mm}
\caption{
The cumulative in-situ fraction, defined as the fraction of all stars that formed prior to a given redshift that formed within the most massive progenitor (MMP).
Thin colored lines show each simulation, including 6 isolated hosts (solid) and 6 LG-like hosts (dotted).
We also show the median across all simulations (thick black line), across isolated hosts (solid grey line), and across LG-like hosts (dashed grey line).
The horizontal black dotted line at 0.5 shows our definition for progenitor `formation', above which the host galaxy has transitioned to having formed the majority of its stars in-situ.
Selecting all stars within 15 kpc at $z = 0$, the median progenitor formation across all 12 hosts occurred at $z = 3.5$ (12.0 Gyr ago).
For stars in the inner/bulge region, $d_2$, the main progenitor emerged earlier, at $z = 5.2$ (12.6 Gyr ago).
Reflective of Fig.~\ref{fig:mmp}, for $d_{15}$, progenitor formation occurred earlier for LG-like hosts ($z = 4.9$, 12.5 Gyr ago) than isolated hosts ($z = 2.7$, 11.3 Gyr ago).
For $d_2$, isolated host progenitors formed at $z = 3.9$ (12.2 Gyr ago), while LG-like hosts were never below 0.5, that is, the majority of stars at all ages in their bulge region today formed in-situ in the MMP.
\textit{Thus, main progenitors of LG-like paired hosts formed/emerged significantly earlier.}
}
\label{fig:insitu}
\end{figure*}

Having explored the ex-situ fraction as a function of progenitor mass, we next examine the total in-situ fraction for each host.
We aim to understand the transition from (1) an early period when most stars formed ex-situ across several progenitors that (eventually) merge together, to (2) a later period when a single main progenitor dominated the stellar mass growth via in-situ star formation.
Thus, we define `in-situ' star formation as that occurring in the MMP, and we define main progenitor `formation' (or `emergence') as the transition from (1) to (2).
This is different from other definitions of progenitor `formation', such as when the galaxy formed a certain percentage of its current $\Mstar$ (see Section~\ref{sec:mmp growth}).
Our goal is instead to quantify when a single main progenitor dominated the stellar mass assembly.
To calculate the fraction of in-situ star formation, we select stars within the host-centric distance selections at $z = 0$ that are older than a given redshift.
Of these, we then select the stars that formed inside of the MMP, and divide this stellar mass by the total stellar mass at $z = 0$ that is older than this redshift.
Finally, we cumulatively sum these fractions to get the $\rm f_{in-situ, cumul}$, shown in Fig.~\ref{fig:insitu}.

Fig.~\ref{fig:insitu} shows this cumulative in-situ fraction as a function of redshift for the $d_{15}$ and $d_2$ distance selections.
We show each simulation as a thin colored line (solid for isolated hosts, dotted for LG-like hosts), along with medians for the total, isolated, and LG-like hosts.
The horizontal dotted line shows a cumulative in-situ fraction of 0.5, which we use to define the formation of a single `main' progenitor.

Across all 12 hosts, the median formation of the whole galaxy (left panel) occurred at $z \sim 3.5$ (12.0 Gyr ago), and at this redshift, the MMP was $\sim 1$ per cent of its present $\Mstar$ (see bottom left panel in Fig.~\ref{fig:mmp}) and the MMP halo was $\sim 10$ per cent of its present $\Mthm$ (Fig.~\ref{fig:mmpdm}).
Formation in LG-like and isolated hosts occurred around $z = 4.9$ (12.5 Gyr ago) and $z = 2.7$ (11.3 Gyr ago) respectively.
However, both host types converged to similar in-situ fractions by $z \lesssim 1$ (7.8 Gyr ago), so these environmental differences were important only in early formation.
The median cumulative in-situ fraction at $z = 0$ is $\sim 93$ per cent, which is consistent with previous results in \citet{AnglesAlcazar17}, who analyzed the FIRE-1 simulations (including a version of m12i) and found that in-situ star formation dominates $\geq 95$ per cent of the stellar mass growth at MW masses.
The NIHAO simulations also show similar in-situ fractions \citep{NIHAO}.

Consistent with (radial) inside-out formation, the inner bulge region of the host (right panel) established itself in a single main progenitor earlier than the overall galaxy, with median formation at $z \sim 5.2$ (12.6 Gyr ago).
Interestingly, the median for LG-like hosts never extended below 0.5, at least back to $z = 6$ (12.8 Gyr ago), meaning that these stars formed in a single main progenitor \textit{at all redshifts that we probe}.
By contrast, bulge stars in isolated hosts formed in a single main progenitor only at $z \lesssim 3.9$ (12.2 Gyr ago).

While we have presented the \textit{cumulative} in-situ fraction, considering all stars that formed to a given redshift, we also examined the \textit{instantaneous} in-situ fraction, using stars that formed within a narrow bin of redshift (not shown).
While this is a more time variable/stochastic metric, we found similar trends overall.
The key difference is that, because the in-situ fraction rises with decreasing redshift, the cumulative value, being an integral quantity, is smaller than the instantaneous value at a given redshift.
Thus, a galaxy transitions above 0.5 at a smaller redshift (typically $\Delta z \sim 0.25$, $\Delta t \sim 0.12$ Gyr) when considering the cumulative in-situ fraction.
Both fractions show the same dependence on host-centric distance selection and the same trend that LG-like hosts formed earlier.
We also note that the in-situ fraction for the host `disk' selection, mentioned in Section~\ref{sec:select}, is similar to $d_{15}$, with comparable formation times.

Finally, while we discuss median trends above, we emphasize \textit{significant} host-to-host scatter in Fig.~\ref{fig:insitu}.
For example, for the $d_{15}$ selection, Romeo never had an in-situ fraction below 0.5 back to $z = 6$ (12.8 Gyr ago), while Thelma become dominated by in-situ mass growth only at $z < 1.7$ (9.9 Gyr ago).
We also note that the in-situ fractions for some hosts can temporarily decline with time.
For instance, in m12c the in-situ fraction decreased from $z \sim 4$ to 2 (12.2 Gyr to 10.4 Gyr ago), which indicates enhanced star formation in progenitors other than the MMP.

Overall, the general trends for both host-centric distance selections are (1) LG-like hosts formed earlier than isolated hosts, (2) the median and scatter tend to converge at $z \lesssim 1$ (7.8 Gyr ago), and (3) no single progenitor forms later than $z \sim 1.7$ (9.9 Gyr ago).

\subsection{When did a dominant-mass progenitor emerge?}
\label{sec:mrs}

\begin{figure*}
\centering
\begin{tabular}{c @{\hspace{-0.5ex}} c}
\includegraphics[width=0.45\linewidth]{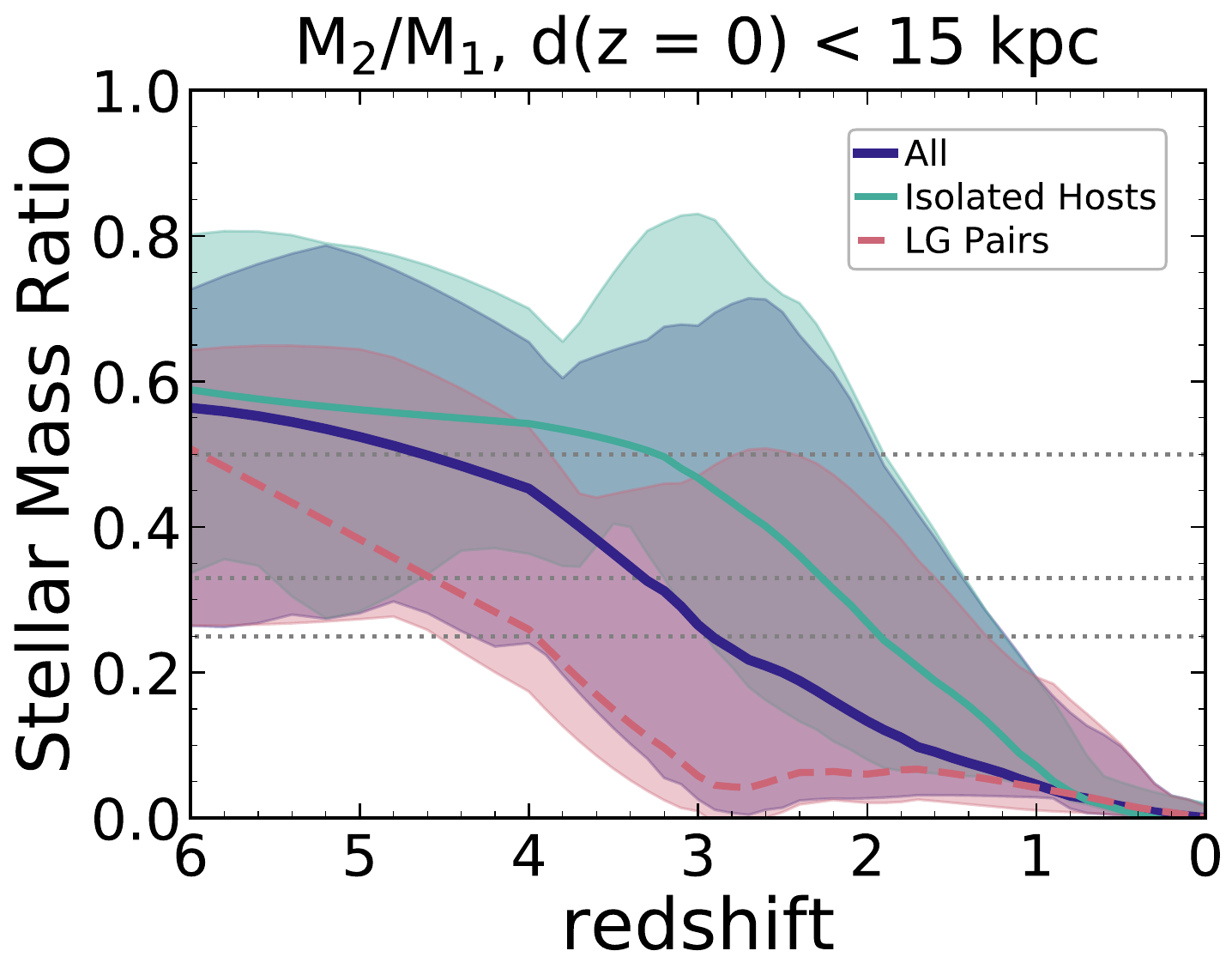}&
\includegraphics[width=0.45\linewidth]{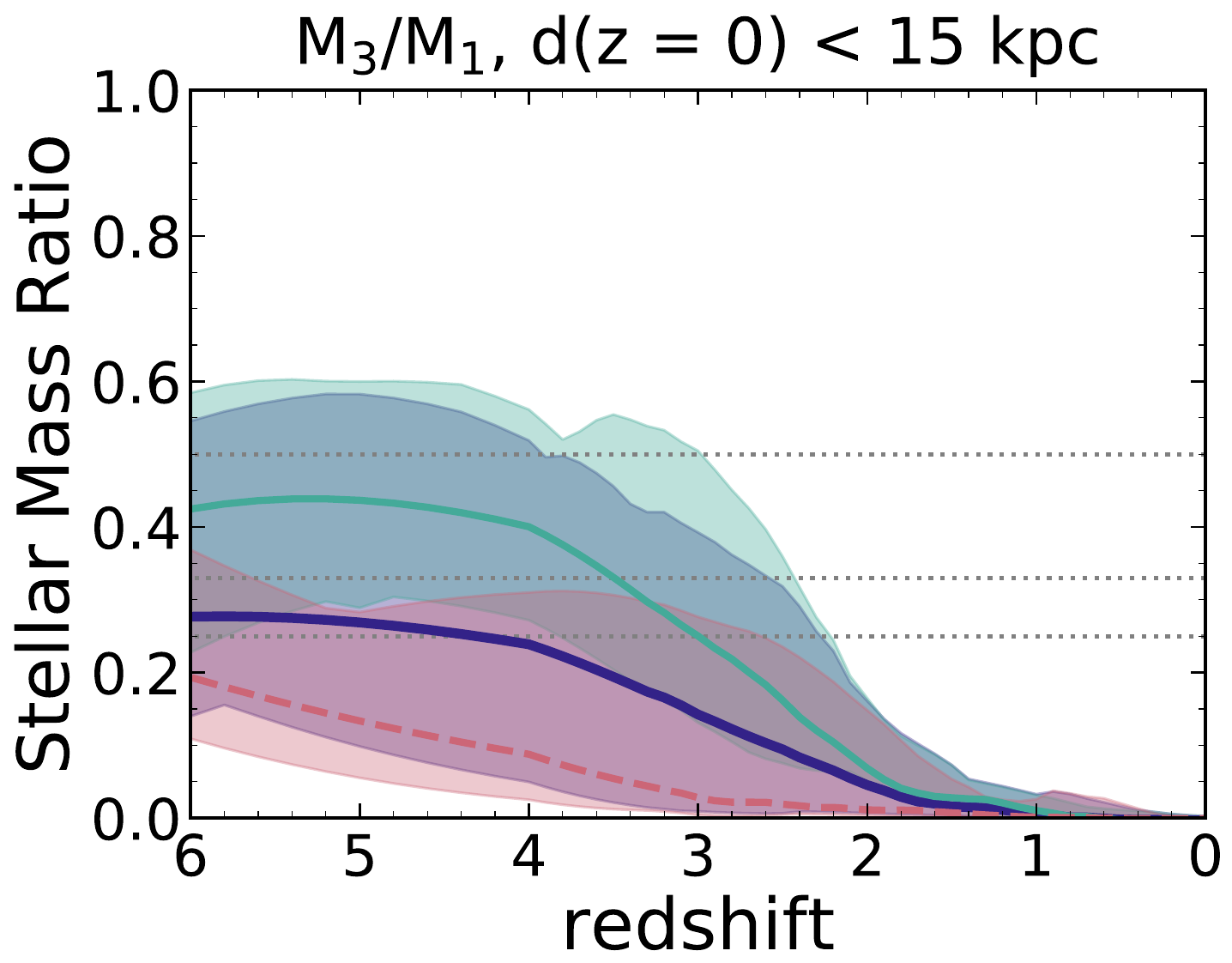}
\end{tabular}
\vspace{-2 mm}
\caption{
For all progenitor galaxies that contribute stars to host-centric $d_{15}$, the ratio of the stellar mass of the second ($\rm M_2$, left) or third ($\rm M_3$, right) most massive galaxy to that of the most massive progenitor (MMP, $\rm M_1$) versus redshift.
We show the (smoothed) median across all 12 hosts (solid purple), isolated hosts (solid green), and LG-like hosts (dashed salmon), with the 68 per cent scatter across all hosts in the shaded regions.
The horizontal dotted lines show 1:2, 1:3, and 1:4 mass ratios for reference.
When each system crosses below these values, the MMP ($\rm M_1$) increasingly becomes the dominant galaxy across all progenitors.
Across all 12 hosts, this transition occurred at $z = 4.6$ (12.5 Gyr ago), $z = 3.3$ (11.8 Gyr ago) and $z = 2.9$ (11.5 Gyr ago) for a 1:2, 1:3, and 1:4 ratio in $\rm M_2/M_1$.
$\rm M_3$ is rarely within a factor of 2 of $\rm M_1$, and $\rm M_3/M_1$ drops below 1:4 at $z = 4.2$ (12.3 Gyr ago).
Again, the MMP of LG-like hosts becomes dominant earlier than for isolated hosts: $\rm M_2/M_1$ for LG-like hosts crossed below 1:2, 1:3, and 1:4 at $z = 6$ (12.8 Gyr ago), $z = 4.6$ (12.5 Gyr ago), and $z = 3.9$ (12.2 Gyr ago), while for isolated hosts these transitions occurred at $z = 3.3$ (11.8 Gyr ago), $z = 2.3$ (10.9 Gyr ago), and $z = 1.9$ (10.2 Gyr ago).
}
\label{fig:mrs}
\end{figure*}

Having examined the in-situ fraction of star formation, we also examine a second metric of main progenitor `formation': when the (instantaneous) stellar mass of the single MMP galaxy dominated that of any other progenitor.
Fig.~\ref{fig:mrs} shows the ratios of the second- and third-most massive progenitors relative to the MMP, $\rm M_2/M_1$ and $\rm M_3/M_1$ (left and right panels respectively), as a function of redshift, for progenitors that contribute stars to $d_{15}$.
We show the median across all 12 hosts (solid purple), 6 isolated hosts (solid green), and 6 LG-like hosts (dashed salmon), as well as their respective 68 per cent scatters via the shaded regions.
Fig.~\ref{fig:mrs} also shows stellar mass ratios of 1:2, 1:3, and 1:4 via horizontal dotted black lines.

We find the same qualitative trends in both panels, so we focus primarily on $\rm M_2/M_1$.
However, we note that the median $\rm M_3/M_1$ ratios were rarely above 1:4 across the total sample, and transitioned below this near $z \sim 4.2$ (12.3 Gyr ago).
The (higher) median for the isolated hosts never reached 1:2, and it transitioned below 1:4 later, around $z \sim 3$ (11.6 Gyr ago).

Focusing on the $\rm M_2/M_1$ ratio (left), we again find that LG-like hosts formed earlier than isolated hosts.
Using 1:3 as a fiducial mass ratio, the redshifts where the medians crossed below 1:3 were $z = 4.6$ (12.5 Gyr ago) and 2.3 (10.9 Gyr ago) for the LG-like and isolated hosts, respectively, which is consistent with values in the previous section for $d_{15}$.
If we examine all simulations, this transition occurred between the two at $z = 3.3$ (11.8 Gyr ago).
Again, the MMP was $\sim 1$ per cent of its present $\Mstar$ at these redshifts, meaning that \textit{the main progenitor formed/emerged as the dominant galaxy well before it formed most of its current mass}.
For context, the present ratio of $\Mstar$ between the LMC and the MW is about 1:30 ($\sim 0.033$) and for M33 and M31 it is roughly 1:20, or $\sim 0.047$ \citep[see compilation in ][]{GarrisonKimmel19b}.

While we focus on the median trends, we again emphasize the significant scatter in formation histories.
We also checked our results using selection of both $d(z = 0) < 300$ and $< 2 \kpc$ and saw little  change, because the MMP contributes stars throughout the host galaxy at $z = 0$.

\subsection{When did the main progenitor form?}
\label{sec:reds}

\begin{figure*}
\centering
\begin{tabular}{c}
    \includegraphics[width=0.65\linewidth]{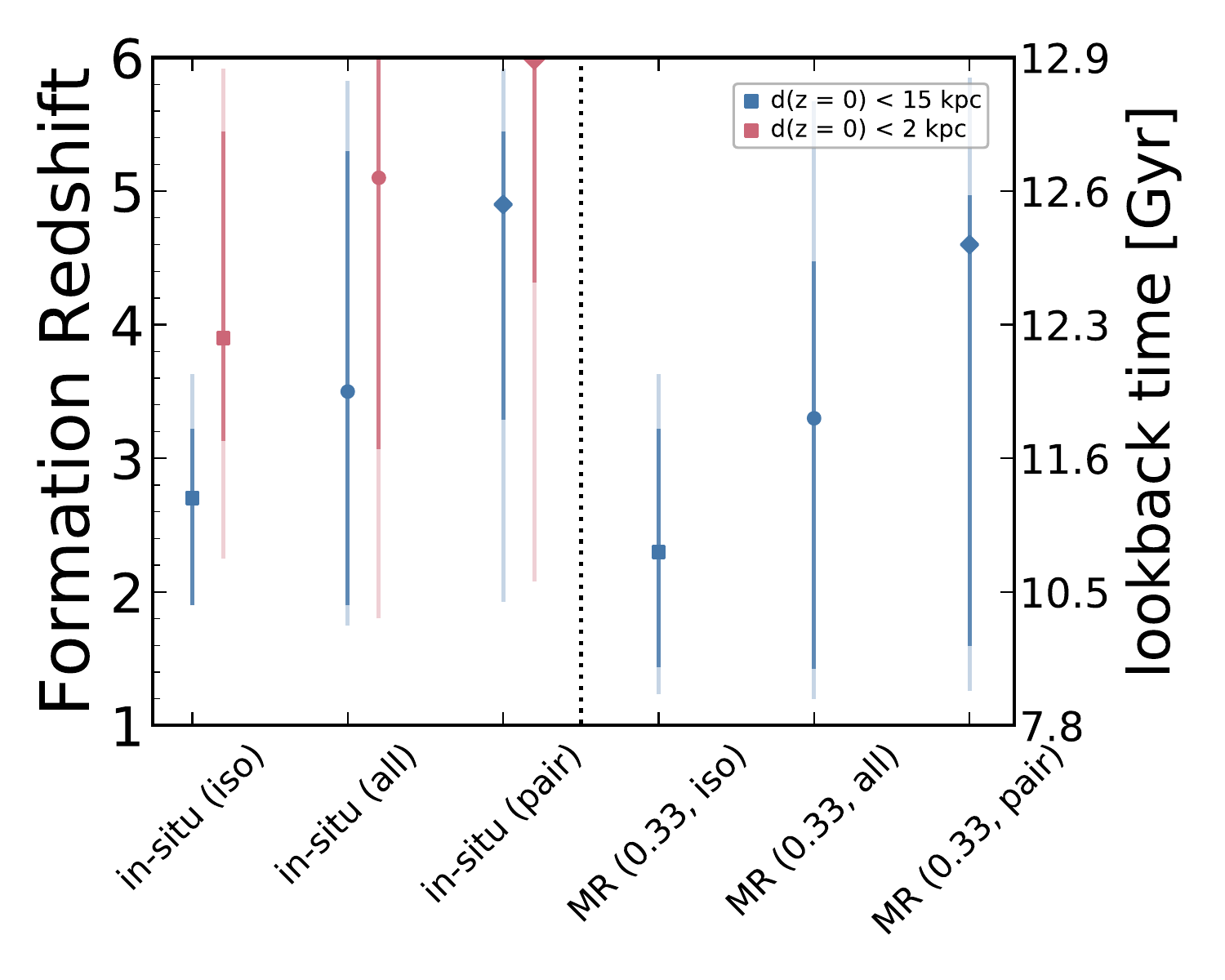}
\end{tabular}
\vspace{-2 mm}
\caption{
Summary of the `formation' redshifts (left axis) or lookback times (right axis) of the main progenitors of our simulated MW/M31-mass galaxies.
The left half shows formation defined when the cumulative fraction of stars that formed in-situ exceeds 0.5 (see  Section~\ref{sec:insitu}), selecting stars at two host-centric distances at $z = 0$, corresponding to the galaxy + inner stellar halo, $d_{15}$ (blue points), and the inner bulge region, $d_2$ (red points).
The right half shows formation defined when the most massive progenitor exceeds a 3:1 stellar mass ratio with respect to the second most massive progenitor (see Section~\ref{sec:mrs}).
Points show the median across the sample and vertical bars show the 68 per cent (darker) and 95 per cent (lighter) scatters, using all 12 hosts (circles), only 6 isolated hosts (squares), and only 6 LG-like hosts (diamonds).
For the LG-like hosts at $d_2$, we show the in-situ formation redshift at $z = 6$ (12.8 Gyr ago) as a lower limit.
Considering the entire host galaxy, the main progenitor formation times are similar for in-situ and 3:1 mass-ratio metrics, being $z \sim 3.4$ ($11.9$ Gyr ago) for the full sample, though significantly later at $z \sim 2.5$ (11.1 Gyr ago) for isolated hosts and earlier at $z \sim 4.7$ (12.5 Gyr ago) for the LG-like paired hosts, with \textit{significant} scatter across different hosts.
}
\label{fig:reds}
\end{figure*}

Here, we present the redshifts corresponding to our definitions of progenitor formation regarding the two metrics in Sections~\ref{sec:insitu} and \ref{sec:mrs}: the cumulative in-situ star formation fraction and the $\rm M_2 / M_1$ (instantaneous) $\Mstar$ ratio.
For the in-situ fractions, we use 0.5 as our fiducial threshold to determine when the progenitor formed.
Regarding the mass ratio, we choose 1:3 to define formation; this metric does not depend on host-centric distance selection.
Fig.~\ref{fig:reds} shows the total median, as well as the medians of isolated and LG-like hosts separately, with the 68 per cent and 95 per cent scatter in the dark and light vertical bars respectively.

Considering main progenitor formation based on in-situ star formation for $d_{15}$, the median formation for all 12 hosts, isolated hosts, and LG-like hosts occurred at $z = 3.5$ (12.0 Gyr ago), 2.7 (11.3 Gyr ago), and 4.9 (12.5 Gyr ago) respectively, but the scatters cover a wide range of $z \sim 1.7 - 6$ ($9.9 - 12.8$ Gyr ago).
For $d_2$, because the median for LG-like hosts never extended below 0.5 at $z < 6$, we set their formation redshift to be $z = 6$ as a lower limit, making the total, isolated, and LG-like host medians $z = 5.1$ (12.6 Gyr ago), 3.9 (12.2 Gyr ago), and 6 (12.8 Gyr ago), with similar redshift scatter.
LG-like hosts had earlier formation times, and we see the same trend of earlier formation for smaller host-centric distance cuts.

If we instead examine main progenitor formation based on instantaneous stellar mass ratio, it reached 1:3 at $z = 3.3$ (11.8 Gyr ago), 2.3 (10.9 Gyr ago), and 4.6 (12.5 Gyr ago) for the total sample, isolated hosts, and LG-like hosts, respectively, with scatter of $z \sim 1.2 - 5.8$ ($8.6 - 12.8$ Gyr ago).
Again, LG-like hosts formed earlier, we find similar median formation times compared with in-situ based formation.

\subsection{Does the mass growth of the dark-matter halo depend on environment?}
\label{sec:mmphalo}

\begin{figure}
\centering
\includegraphics[width = 0.99 \columnwidth]{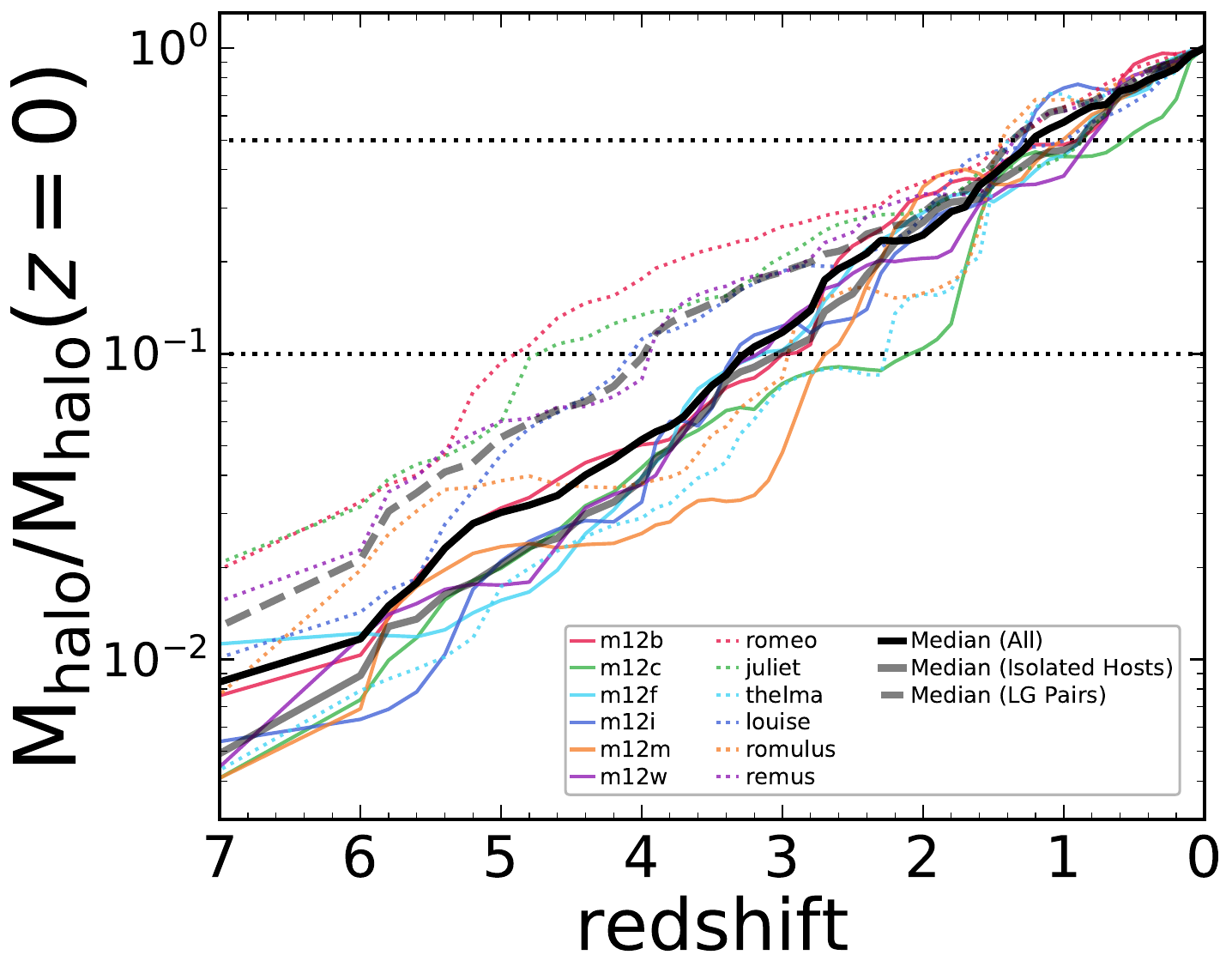}
\vspace{-3 mm}
\caption{
Similar to Fig.~\ref{fig:mmp} (bottom left), but showing the \textit{dark-matter halo} mass of the most massive progenitor (MMP) of each MW/M31-mass host, normalized to each host's $\Mthm$ at $z = 0$, as a function of redshift (bottom axis) or lookback time (top axis).
We show the 6 isolated hosts (solid) and 6 LG-like paired hosts (dotted) in thin colored lines, the median across all 12 hosts (thick solid black), and the medians for isolated (thick solid grey) and LG-like (thick dashed grey) hosts.
The dotted horizontal lines show 10 per cent and 50 per cent of the final mass.
For the total median, the MMP halo reached 10 per cent of its final mass at $z \sim 3.3$ (11.8 Gyr ago).
However, isolated hosts reached this later at $z \sim 3$ (11.6 Gyr ago), while LG-like hosts reached it earlier at $z \sim 4$ (12.2 Gyr ago).
At later times ($z \lesssim 2$), the mass growth histories of isolated and LG-like hosts are more similar, so this environmental effect is weaker for later-term growth.
We conclude that this more rapid \textit{halo} mass growth in denser proto-LG-like environments at early times likely drives the enhanced \textit{stellar} mass growth in Fig.~\ref{fig:mmp}.
}
\label{fig:mmpdm}
\end{figure}

Finally, we seek to understand more deeply why the most massive progenitor (MMP) of a LG-like host experiences more rapid stellar mass growth (Fig.~\ref{fig:mmp}) and earlier `formation' of a main progenitor (Fig.~\ref{fig:insitu}) than isolated hosts.
Specifically, we investigate whether the mass growth of the DM halo reflects these trends as well.
We proceed as with Fig.~\ref{fig:mmp}, but instead we measure the DM halo mass of the MMP galaxy.
For most progenitors, the highest-mass galaxy resides in the highest-mass halo.
However, because of scatter in stellar versus halo mass growth, especially at early times when there was no clear main progenitor, the MMP galaxy might not reside in the MMP halo.
In these cases, we select the halo with the highest DM halo mass as the MMP, but we find nearly identical results if instead we show the DM halo mass of the most massive galaxy.

As Fig.~\ref{fig:mmpdm} shows, we find qualitatively similar results for halo mass growth as for stellar mass growth: LG-like hosts grow in mass more rapidly than isolated hosts.
Specifically, across all 12 hosts, the MMP halo reached 10 per cent of its final mass by $z \sim 3.3$ (11.8 Gyr ago), but isolated hosts reached this later at $z \sim 3$ (11.6 Gyr ago) and LG-like hosts reached it earlier at $z \sim 4$ (12.2 Gyr ago), with $\Delta z \sim 1$ and $\Delta t \sim 0.6$ Gyr.
At later times ($z \lesssim 2$), we find some enhanced growth for LG-like hosts, but the mass growth histories are more similar, so this environmental effect is weaker for later-term halo growth.
These median redshifts are all \textit{earlier} than those for stellar mass in Fig.~\ref{fig:mmp}, given that central gravitational potential of a DM halo establishes itself earlier than galaxy that it hosts.

We thus conclude that these differences in halo mass growth likely cause (at least to first order) the differences in stellar mass growth and satellite populations, especially because these environmental differences persist back to the initial collapse of theses halos at $z \gtrsim 7$.
This is perhaps not surprising, given that halos in LG-like environments formed in denser regions that should collapse earlier than isolated halos \citep{Gallart15}.

Using a larger sample of 24 paired and 24 isolated host halos in the ELVIS DMO suite of simulations, \cite{GarrisonKimmel14} did not find major differences in the median formation times of LG-like halos compared with isolated halos.
Similarly, a study by \citet{ForeroRomero11} using DMO simulations from both the Constrained Local Universe Simulations (CLUES) project and Bolshoi \citep{Riebe13}, did not see differences in the formation times of isolated versus LG-like hosts.
However, both of these works measured halo `formation' based on the redshift when a halo formed 50 per cent of its final mass at $z = 0$.
\citet{GarrisonKimmel14} found that both isolated and LG-like hosts had $z_{\rm form} \sim 1.1$, and we find nearly identical results for our baryonic simulations.
The key difference in Fig.~\ref{fig:mmpdm} is the \textit{early} formation history of the DM halo, which appears to affect the stellar mass growth to even lower redshifts, as evidenced in Fig.~\ref{fig:mmp}, where stellar mass growth histories of LG-like versus isolated hosts diverge at $z > 0$.

As mentioned in Section~\ref{sec:sims}, the simulations assume flat $\Lambda$CDM cosmology, and the 6 cosmological parameters span a range of values that are consistent with \citet{Planck18}.
However, not all simulations used the same cosmology, and one may wonder if this affects their formation times.
These differences do not appear to correlate strongly with halo formation time.
The Latte suite, Thelma \& Louise, and Romulus \& Remus adopt the most similar cosmologies.
The most distinct cosmology, for Romeo \& Juliet and m12w, includes both LG-like hosts and an isolated host, and although Romeo \& Juliet did form the earliest (along with Romulus, with $z \sim 6$ for Romeo, $z \sim 5.3$ for Juliet and Romulus), m12w has a relatively late formation time ($z \sim 2.7$).
Furthermore, Thelma \& Louise span almost the entire range, with Thelma being the latest forming of all hosts and Louise being one of the earliest.
Thus we conclude that environment, and not slight differences in cosmology, is the primary cause of the difference in halo/galaxy formation history.

\section{Summary and Discussion}
\label{sec:disc}

\subsection{Summary}
\label{sec:sum}

Using a suite of 12 FIRE-2 cosmological zoom-in simulations of MW/M31-mass galaxies, we explored their formation histories, to understand when a single main progenitor formed/emerged and quantify the hierarchical build-up from a progenitor population.
We defined main progenitor formation in two ways: (1) when the growth of the MMP transitions from mostly ex-situ to in-situ star formation, and (2) by mass dominance (3:1 ratio or closer) of the MMP compared to other progenitors.
The questions that we posed in the introduction and our corresponding answers are:
\renewcommand{\labelenumi}{\alph{enumi})}
\renewcommand{\labelenumii}{\roman{enumii})}
\begin{enumerate}
\item \textit{What were the building blocks (progenitor galaxies) of MW/M31-mass galaxies, and how many were there across cosmic time?}
    \begin{enumerate}
    \item About 100 progenitor galaxies with $\Mstar \geq 10^5 \Msun$, $\sim 10$ with $\Mstar \geq 10^7 \Msun$, and $\sim 1$ with $\Mstar \geq 10^9 \Msun$ formed a typical MW/M31-mass system. Thus, there were $\sim 5$ times as many dwarf-galaxy progenitors with $\Mstar > 10^5 \Msun$ at $z \sim 4 - 6$ ($12.2 - 12.8 \Gyr$ ago) than survive to $z = 0$ (Fig.~\ref{fig:mfs}).
    \item The slope of the progenitor galaxy mass function was steeper with increasing redshift, which qualitatively agrees with observational and simulation results regarding the overall galaxy population (Fig.~\ref{fig:mfs}).
    \item At all redshifts, the ex-situ stellar mass of the accreted population was dominated by the few most massive progenitors, and the ex-situ fraction monotonically increased with redshift (Fig.~\ref{fig:wmfs}).
    \end{enumerate}

\item \textit{When did the main progenitor of a MW/M31-mass galaxy form/emerge?}
    \begin{enumerate}
    \item Across all 12 hosts, a single main progenitor typically formed/emerged around $z \approx 3.3 - 3.5$ (11.8 - 12.0 Gyr ago) (Figures~\ref{fig:insitu}, \ref{fig:mrs}, \ref{fig:reds}).
    \item Stars in the inner bulge region formed in a single main progenitor earlier, typically at $z \approx 5.2$ (12.6 Gyr ago) across all 12 hosts (Fig.~\ref{fig:insitu}).
    \item Across all 12 hosts, the MMP reached 10 per cent of its present stellar mass by $z = 1.7$ (9.9 Gyr ago) and 50 per cent by $z = 0.5$ (5.1 Gyr ago). Thus, a single main progenitor typically formed/emerged when the host had only a few percent of its final stellar mass (Fig.~\ref{fig:mmp}).
    \end{enumerate}

\item \textit{Does the formation of MW/M31-mass galaxies depend on their environment, specifically, comparing isolated hosts to those in LG-like pairs?}
    \begin{enumerate}
    \item LG-like hosts reached 10 per cent and 50 per cent of their present stellar mass around $z = 2.4$ (11.0 Gyr ago) and $z = 0.8$ (6.9 Gyr ago), respectively.
    This was significantly earlier than when isolated hosts reached the same fractional masses: $z = 1.5$ (9.4 Gyr ago) for 10 per cent and $z = 0.5$ (5.1 Gyr ago) for 50 per cent (Fig.~\ref{fig:mmp}).
    \item Similarly, a single main progenitor of a typical LG-like paired host formed  significantly earlier ($z_{\rm form} = 4.6 - 4.9$, $\sim 12.5$ Gyr ago) than for a typical isolated host ($z_{\rm form} = 2.3 - 2.7$, 10.9 - 11.3 Gyr ago) (Figs.~\ref{fig:insitu}, \ref{fig:mrs}, \ref{fig:reds}).
    This is likely because their DM halos formed earlier (Fig.~\ref{fig:mmpdm}).
    \item We find weaker differences between the overall progenitor galaxy populations for LG-like versus isolated hosts across time: the primary difference is that the number of progenitors peaked later for isolated hosts, reflecting their overall later formation histories (Fig.~\ref{fig:mfs}).
    \end{enumerate}
\end{enumerate}

\subsection{Discussion}
\label{sec:disc2}

Our simulations show that LG-like host galaxies were more massive than isolated hosts (of the same mass at $z \sim 0$), before $z \sim 2$, back to at least $z = 7$.
This result is consistent with a related SFH-based analysis of the same simulations in \citet{GarrisonKimmel19a}.
As we showed in Section~\ref{sec:mmphalo}, this difference is reflected in the early formation histories of the DM host halos, and it may be exacerbated in stellar mass growth if earlier halo formation promotes more metal production throughout the proto-volume, which would make gas cooling more efficient \citep{GarrisonKimmel19a}.
That paper also found no difference in the SFHs of the satellite galaxies of FIRE-2 isolated versus LG-like hosts, but they \textit{did} find differences in the formation times of central dwarf galaxies in the `near-field' around LG-like versus isolated hosts, a population that we did not examine in this work. 

Our results have key implications for studies of the early Universe and cosmic reionization. 
The slope of the galaxy luminosity (and mass) function at the faint (low-mass) end informs the contribution of low-mass galaxies to the ionizing flux during cosmic reionization at $z \gtrsim 7$.
For example, if one naively extrapolates the slope to arbitrarily low mass, galaxies with ultra-violet (UV) luminosity $M_{UV} \gtrsim -10$ generated most of the ionizing photons during reionization \citep[$\sim 50 - 80$ per cent;][]{Weisz17}.
Thus, the evolution of the galaxy luminosity/mass function is of considerable interest.
Several observational and theoretical works indicate that the slope of the faint end of the galaxy luminosity/mass function steepens with redshift \citep[e.g.][]{Bouwens15, Song16}.
For example, \citet{Bouwens15} show that the faint-end slope of the UV luminosity function evolves from $\alpha \sim -1.64$ at $z \sim 4$ to $\alpha \sim -2.02$ at $z \sim 8$.
\citet{Ma18} show that in FIRE-2 simulations of larger populations of galaxies, the slope of the low-mass end of the stellar mass function decreases from $\alpha \sim -1.8$ at $z = 6$ to $\alpha \sim -2.13$ at $z = 12$.
\citet{Graus16} summarized several observational works and applied abundance matching to the DMO ELVIS simulations in order to calculate galaxy stellar mass and luminosity functions, finding a slight steepening of faint-end slope from $z = 2$ to $z = 5$.

It is not a priori obvious that the mass function of the progenitors of MW/M31-mass galaxies, which represent a biased region of all galaxies, reflects the overall galaxy population at a given redshift.
This is an important question, because the faintest galaxies at $z \sim 7$ will be too faint even for direct JWST observations.
Recent works have proposed using resolved stellar populations and SFHs of dwarf galaxies in the LG to infer the faint-end slope of the UV luminosity function at high redshifts, which already provides evidence for a break/rollover in the faint-end slope of the UV luminosity function at $z \sim 7$, given the number of ultra-faint dwarf (UFD) galaxies in the LG \citep{BoylanKolchin14, BoylanKolchin15, Weisz17}.
On the one hand, our results in Fig.~\ref{fig:mfs} (right) show that the progenitors of MW/M31-mass systems do show a similar steeping of mass-function slope to higher redshifts, which is at least qualitatively consistent with the overall galaxy population.
However, our results also show that a significant fraction ($\sim 80$ per cent) of these progenitor dwarf galaxies have disrupted into the MW, which means that any inference from the LG population today is missing such progenitors in the early Universe.
Thus, our results suggest that this `near-far' approach remains promising, but more work is needed to explore quantitatively how representative the proto-LG environments were at high redshifts, which we plan to pursue in future work.

Wide-field surveys currently are measuring elemental abundances, ages, and phase-space distributions for millions of stars throughout all components of the MW, and their sampling rate is expected to continue to increase in the coming decade.
A principle aim of these surveys is to reconstruct the formation history of the MW.
Our analysis shows that MW/M31-like galaxies were assembled from $\sim 100$ distinct dwarf galaxies with $\Mstar \geq 10^5 \Msun$, a majority of which merged by $z \sim 2$ (Fig.~\ref{fig:mfs}).
Each of these dwarf galaxies had a unique orbit, and some likely had distinct elemental abundance patterns.
In principle, it may be possible to identify members of many distinct progenitors by identifying them as clumps in a high-dimensional chemo-dynamic space \citep[e.g.][]{Ting15}, even in the Solar neighborhood \citep[e.g. see recent work by][]{Necib19}.
Achieving this in practice is challenging because progenitors that merged prior to $z \sim 3$ are likely thoroughly phase-mixed by $z = 0$ \citep[see][]{ElBadry18}.
Such phase-mixing of the earliest-accreted progenitors occurs naturally during merging, and is likely exacerbated by stellar feedback-driven oscillations of the gravitational potential at early times \citep[e.g.][]{ElBadry16}.

On the other hand, progenitors accreted later ($z \lesssim 2$) -- particularly the most massive ones -- likely still can be identified.
Indeed, there is already compelling evidence that much of the MW's inner stellar halo was formed by a single massive progenitor, whose stars remain dynamically coherent and chemically distinguishable \citep[e.g.][]{Helmi18, Belokurov18}.
This fact is not surprising in the context of our simulations: because the progenitor mass functions of MW/M31-mass galaxies are relatively shallow (Fig.~\ref{fig:mfs}), our simulations generically predict that for any given formation redshift, most of the mass in the stellar halo was contributed by the few most massive progenitors (Fig.~\ref{fig:wmfs}).
We finally note that this has also been seen in MW/M31-mass galaxies from the Illustris simulations \citep[e.g.][]{Dsouza18}.

Populations of stars that are both old \textit{and} metal-poor tend to be more centrally concentrated despite the fact that the individual fractions of old \textit{or} metal-poor stars increases with radius from the galactic center \citep{Starkenburg17, ElBadry18}.
But, because of continued star formation, central regions tend to get crowded over time and the fraction of old/metal-poor stars to the total stellar population becomes vanishingly small.
Current stellar surveys (e.g. RAVE, GALAH, APOGEE) thus have a better chance of detecting these stars outside of the solar circle (> 8 kpc) and have already found a large number with [Fe/H] < -2.
Despite this fact, APOGEE has observed $\sim 5100$ stars near the Galactic bulge and have found a subset of these stars believed to be both old, but more metal-rich ([Fe/H] $\sim -1$, \cite{Schiavon17}).
For future work, both within the MW and beyond, LSST also will be capable of finding RR Lyrae stars, which tend to be old ($> 10 \Gyr$), within the LG volume \citep{Oluseyi12}.

The formation times that we obtain for the entire galaxy (inner stellar halo + disk) are in line with those reported for the halo and thick disk in \citet{Gallart19} ($z \sim 2 - 4.2$, $\sim 10.5 - 12.3$ Gyr ago).
By analyzing the difference in elemental abundances, they determined that the merger between Gaia-Enceladus/Sausage and the MW progenitor was about a 1:4 ratio, and took place roughly 10 Gyr ago, agreeing with previous work \citep[e.g.][]{Helmi18, Belokurov18, Nogueras-Lara19}, which would correspond to when the main galaxy was $\sim 10$ per cent of its total stellar mass, or $\sim 30$ per cent of its total halo mass.
Interestingly, around 10 Gyr the $\rm M_2 / M_1$ ratios in the simulations are more pronounced, at a 1:10 ratio for the total sample, but closer for the isolated hosts, $\sim$ 1:4.
Given that the work in \citet{Gallart19} was done for a LG-like host (the MW itself), assuming Gaia-Enceladus/Sausage was the second most massive component in the system at that time, our results suggest the ratio should be closer to 1:10 or 1:20.
At these times, the thick disk was already in place, so this accretion event could have dynamically heated some of these stars into the halo, as well as provide a fresh reservoir of gas for further star formation.

\citet{Nogueras-Lara19} estimated that the nuclear disk, embedded within the bulge, must have formed over 80 per cent of its stars more than 8 Gyr ago.
They also claim that no significant merger $> 5:1$ occurred in the MW within the last $\sim 10 \Gyr$. 
Using CMDs of bulge stars to construct SFHs, other studies place the bulge at $\gtrsim 10$ Gyr old, with no traces of a younger stellar population \citep{Zoccali03, Valenti13, Renzini18, Barbuy18}.
\citet{Bernard18} suggest that 50 per cent of these stars formed before $\sim 10$ Gyr, and 80 per cent formed before $\sim 8$ Gyr.
The median formation times that we find for the inner bulge region are $\sim 2$ Gyr earlier than the 10 Gyr lower-limit, however, the scatter does span a wide range from $\sim 10 - 13$ Gyr.

\citet{Kruijssen19} inferred the MW's assembly history by combining the E-MOSAICS simulation suite, which models globular star cluster formation and evolution in MW/M31-mass galaxies, with the population of observed globular clusters around the MW.
They estimate that the MW formed 50 per cent of its $\Mthm(z = 0)$ around $z = 1.5$ (9.4 Gyr ago) comparable to both our results and that of \citet{GarrisonKimmel14}.
They also suggest that the main progenitor of the MW must have formed half of its stellar mass by $z \sim 1.2$ (8.6 Gyr ago) and that half of the stellar mass in the main galaxy at $z = 0$ formed \textit{across all progenitors} by $z \sim 1.8$ (10.1 Gyr ago).
By selecting globular cluster populations and inferring their evolution through age-metallicity space, \citet{Kruijssen19} also estimated how many mergers of various masses occurred in the MW.
They argue for $\sim 15$ significant mergers throughout the MW's history, with a majority of them ($\sim 9$) happening before $z = 2$.
Although we do not explicitly follow the halo merger tree in our analysis, we see rough consistency in that there were many more progenitor galaxies at high redshift, for all distance selections that we probe.

Although the formation times and mass functions derived in this paper are broadly consistent with other results in the literature, we recognize limitations in our analysis.
First, we examine only 12 MW/M31-mass galaxies.
A larger sample would allow us to probe better the distributions of formation times across a diverse set of formation histories.
We also emphasize again that our galaxies, although comparable to the MW and M31 in many properties, were not created with the intention of exactly reproducing either galaxy.
Therefore, one should not interpret the results in this work as necessarily applying exactly to the MW or M31, but rather as typical cosmological histories for MW/M31-mass galaxies.
Finally, our results on stellar mass assembly are contingent on the accuracy and validity of the  FIRE-2 model.
For example, our assumed meta-galactic UV background from \citet{FaucherGiguere09} causes reionization to occur too early ($z \sim 10$) compared to recent observations \citep[e.g.][$z \sim 7$]{Planck18}; similar results in some other simulation work have reported the same issue \citep[e.g.][]{Onorbe17}.

In the future, an interesting follow-up analysis to our results would be a more thorough investigation into the importance of environment on formation time.
Although we briefly discussed our initial reasoning of this difference in Section~\ref{sec:mmphalo}, more work is needed to provide a comprehensive explanation, including a more rigorous comparison of the merger trees between both isolated and LG-like hosts.
The dependence of galaxy formation time on halo formation is likely only one piece of this story.
Furthermore, there are other important factors that we did not investigate in our analysis, for example, `patchy' reionization that likely did not happen instantaneously or spatially uniformly, as is implemented in FIRE simulations.
Reionization heats gas and suppresses star formation, so a non-uniform reionization would shape a galaxy's SFH differently compared to our simulations \citep[e.g.][]{Lunnan12, Zhu19}.
Patchy reionization also could have important consequences for the metal enrichment of gas and early forming Pop III stars.
These first stars, and the feedback they produce within small, early forming progenitor galaxies, could affect the SFH of the main host at early times \citep[e.g.][]{Koh18, Corlies18}, and a proper implementation of their formation and evolution within simulations is critical.
It also remains unclear exactly how Pop III stars enrich the ISM around them, and how they shape the formation of the subsequent generation of stars within the galaxy.
Finally, the early formation period of galaxies is governed by many low-mass progenitors \citep[e.g.][also see Fig.~\ref{fig:hist}, left column]{Ciardi05}, which are more sensitive to stellar feedback.
It remains unclear how dusty these low-mass progenitors are and how that may affect both star formation and gas heating from the UV background.
Galaxies that are more dusty can both better shield the reionizing photons and more effectively cool gas, which causes earlier star formation of lower mass stars than Pop III.
Because we have not tested any of these additional points, it is unclear which, if any, would play the dominant role in determining a galaxy's formation time at such high redshifts.

We finally note two other studies focused on galactic archaeology within the FIRE simulation collaboration that are investigating the history and evolution of galaxies like the MW.
One example is using chemical tagging of stars to infer where their birth environments were, as well as what other stars they may have co-formed with (Bellardini et al., in prep).
Another involves understanding where a population of metal-poor stars currently in the MW disk came from, as noted in \citet{Sestito19} (Santistevan et al., in prep).


\section*{Acknowledgements}

The authors thank the anonymous referee for their helpful and thorough report.
IBS, AW, and SB received support from NASA, through ATP grant 80NSSC18K1097 and HST grants GO-14734 and AR-15057 from STScI, the Heising-Simons Foundation, and a Hellman Fellowship.
JBH acknowledges support from an ARC Laureate Fellowship.
MBK acknowledges support from NSF grants AST-1517226, AST-1910346, and CAREER grant AST-1752913 and from NASA grants NNX17AG29G and HST-AR-14282, HST-AR-14554, HST-AR-15006, HST-GO-14191, and HST-GO-15658 from the Space Telescope Science Institute, which is operated by AURA, Inc., under NASA contract NAS5-26555. 
CAFG was supported by NSF through grants AST-1517491, AST-1715216, and CAREER award AST-1652522; by NASA through grant 17-ATP17-0067; and by a Cottrell Scholar Award from the Research Corporation for Science Advancement.
We ran and analyzed simulations using XSEDE supported by NSF grant ACI-1548562, Blue Waters via allocation PRAC NSF.1713353 supported by the NSF, and NASA HEC Program through the NAS Division at Ames Research Center.
Some of this work was performed in part at KITP, supported by NSF grant PHY-1748958.

This work also used various python packages including \texttt{numpy} \citep{Numpy}, \texttt{scipy} \citep{SciPy}, and \texttt{matplotlib} \citep{Matplotlib}, as well as NASA's Astrophysics Data System.

\section*{Data availability}
Full simulation snapshots at $z = 0$ are available for m12i, m12f, and m12m at ananke.hub.yt.
The python code used to analyze these data is available at https://bitbucket.org/isantis/progenitor, which uses the publicly available packages https://bitbucket.org/awetzel/gizmo\_analysis, https://bitbucket.org/awetzel/halo\_analysis, and https://bitbucket.org/awetzel/utilities.
Finally, data values in each figures are available at https://ibsantistevan.wixsite.com/mysite/publications.

\bibliographystyle{mnras}
\interlinepenalty=10000
\bibliography{growing_pains}

\bsp	
\label{lastpage}
\end{document}